\documentclass[prb,twocolumn,superscriptaddress]{revtex4-2}
\usepackage[utf8]{inputenc}
\usepackage{amsmath}
\usepackage{physics}
\usepackage{graphicx}
\usepackage{tikz}
\usetikzlibrary{patterns,snakes}
\usepackage{amsfonts}   %for \mathbb{}
\usepackage{comment}
\usepackage{gensymb}
\usepackage{amsbsy,amssymb,amsmath,bm}

\usepackage[sectionbib]{bibunits}
\defaultbibliographystyle{apsrev4-1}
\defaultbibliography{references}

\begin{document}

%\title{Spin-charge separation and resonant valence bond spin liquid in a pyrochlore magnet}
%\title{Spin-charge separation and resonant valence bond spin liquid in a frustrated doped Mott insulator on the pyrochlore lattice}
%\title{Spin-charge separation and resonant valence bond spin liquid in a frustrated doped Mott insulator}
\title{A resonant valence bond spin liquid in the dilute limit of doped frustrated Mott insulators}

\author{Cecilie Glittum}
\affiliation{T.C.M. Group, Cavendish Laboratory, JJ Thomson Avenue, Cambridge CB3 0HE, United Kingdom}
\author{Antonio \v{S}trkalj}
\affiliation{T.C.M. Group, Cavendish Laboratory, JJ Thomson Avenue, Cambridge CB3 0HE, United Kingdom}
\affiliation{Department of Physics, Faculty of Science, University of Zagreb, Bijeni\v{c}ka c. 32, 10000 Zagreb, Croatia}
\author{Dharmalingam Prabhakaran}
\affiliation{Department of Physics, University of Oxford, Clarendon Laboratory, Parks Road, Oxford OX1 3PU, UK}
\author{Paul A. Goddard}
\affiliation{Department of Physics, University of Warwick, Gibbet Hill Road, Coventry, CV4 7AL, UK}
\author{Cristian D. Batista}
\affiliation{Department of Physics and Astronomy, University of Tennessee, 1408 Circle Drive, Knoxville, TN 37996, USA}
\affiliation{Neutron Scattering Division, Oak Ridge National Laboratory, One Bethel Valley Rd, Oak Ridge, TN 37831, USA}
\author{Claudio Castelnovo}
\affiliation{T.C.M. Group, Cavendish Laboratory, JJ Thomson Avenue, Cambridge CB3 0HE, United Kingdom}

\date{\today}

\begin{abstract} % ABSTRACT UNREFERENCED FOR NATURE ARTICLES (150 words guideline)
Ideas about resonant valence bond liquids and spin-charge separation have led to key concepts in physics such as quantum spin liquids, emergent gauge symmetries, topological order, and fractionalisation. Despite extensive efforts to demonstrate the existence of a resonant valence bond phase in the Hubbard model that originally motivated the concept, a definitive realisation has yet to be achieved. Here we present a solution to this long-standing problem by uncovering a resonant valence bond phase exhibiting spin-charge separation in realistic Hamiltonians. We show analytically that this ground state emerges in the dilute-doping limit of a half-filled Mott insulator on corner-sharing tetrahedral lattices with frustrated hopping, in the absence of exchange interactions. We confirm numerically that the results extend to finite exchange interactions, finite-sized systems and finite dopant density. Although much attention has been devoted to the emergence of unconventional states from geometrically frustrated interactions, our work demonstrates that kinetic energy frustration in doped Mott insulators may be essential for stabilising robust, topologically ordered states in real materials.
\end{abstract}

\maketitle
%
%
%%%%%%%%%%%%%%%%%%%%%%%%%%%%%%%%%%%%%%%%%%%%%%%%%%%%%%%%%

\begin{bibunit}

% INTRO WITHOUT HEADING FOR NATURE ARTICLES 
%\section{Introduction
%\label{sec:intro}}
Inspired by Pauling's early theory of resonant covalent-bond-sharing in aromatic molecules, such as benzene, and by Bethe's antiferromagnetic linear chain, Fazekas and Anderson~\cite{Anderson1973,Fazekas1974} proposed in the early 1970s a possible competitor to N\'{e}el order in antiferromagnetically coupled spins: a state where spins are valence bond paired into singlets that resonate quantum mechanically amongst different pairing configurations, thus compensating for the partial loss of antiferromagnetic energy -- forming a resonant valence bond (RVB) state. 
The exciting idea gained further momentum with the discovery of high-temperature superconductivity about a decade later, and with 
the prompt proposal that it may indeed be underpinned by spin-charge separation~\cite{Kivelson1987,Anderson1987,Baskaran1987,Anderson1987b,Zhu1988,Anderson1997} in a system of strongly correlated electrons in an RVB liquid phase 
(for a review, see Refs.~\onlinecite{Anderson2004,Lee2006}). 
The appeal was so great that for decades to follow numerous physicists have been looking for evidence of this suggestion ``{\it along a bewildering variety of routes}''~\cite{Anderson2004}. 

Aside from its relevance to high-temperature superconductivity, the quest for an RVB liquid state led to several fundamental advances, many of which play a central role in our understanding of modern physics. Of particular import, the notion of spin liquids and its relation to deconfinement in gauge theory have been greatly clarified~\cite{Lee2006,Wen2007}, and subsequently extended to further areas of research~\cite{Moessner2001,Balents2002,Kitaev2003,Wen2003}, to encompass new concepts such as topological order and fractionalisation~\cite{Knolle2019}. 

\begin{figure}[ht!]
    \centering
    \includegraphics[width=\columnwidth]{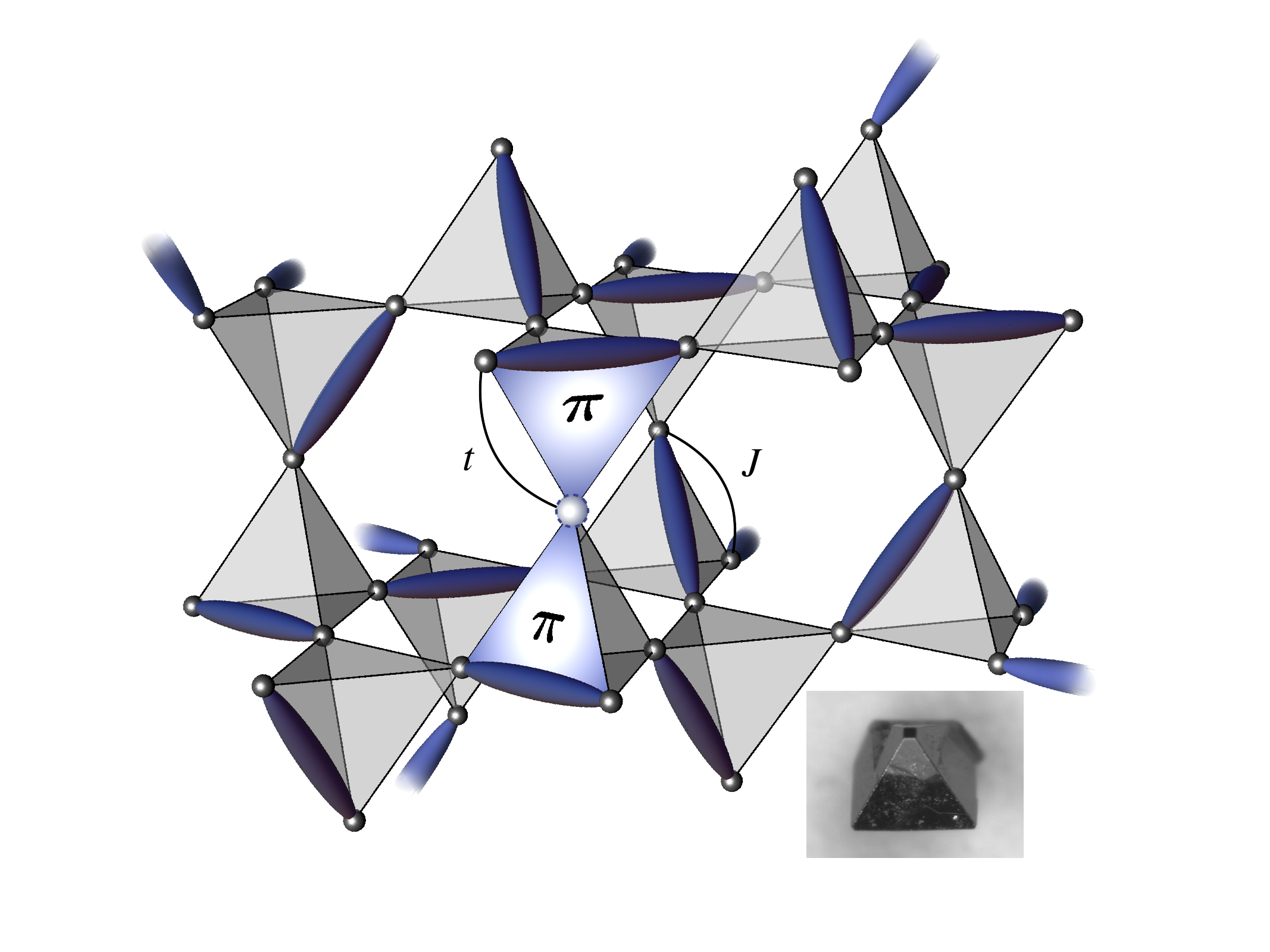}
    \caption{{\it Pictorial illustration of one of the dimer-singlet states}. These states participate in the superposition forming the RVB liquid ground state of the $t$-$J$ model, Eq.~\eqref{eq:ham}, on a lattice of corner-sharing tetrahedra. Singlets are indicated by dark thin ellipsoids, and a holon at a given location by a light sphere. 
    The $\pi$-fluxes depict the phase induced by the singlet coverings (relative to a ferromagnetic background) when the holon moves around the corresponding triangular plaquette. 
    The inset shows an Y$_2$Ir$_2$O$_7$ pyrochlore crystal grown by D.~Prabhakaran in Oxford.}
    \label{fig:pyrochlore lattice}
\end{figure}

Intense work continued throughout the years, yet Anderson's original proposal has hitherto evaded discovery. Following the seminal work by Moessner and Sondhi~\cite{Moessner2001}, which established the possibility of a short-ranged liquid phase in quantum dimer models~\cite{Rokhsar1988}, several efforts attempted to translate this result into a system of SU(2) spins, building on Klein models~\cite{Klein1982}, and effective and decorated dimer models~\cite{Batista2004,Raman2005,Normand2014,Normand2016}. 
It is believed that magnetic frustration in quasi-two-dimensional (quasi-2D) Mott insulators could be responsible for RVB physics; however, parent Hamiltonians that host an exact RVB ground state are often unphysical (see e.g., Ref.~\onlinecite{Schuch2012}). 

In this work, we find a concise and elegant solution -- unintentionally fulfilling the inspirational words of Ref.~\onlinecite{Raman2005}: ``{\it a more important
task perhaps, now that the question of principle is settled, is to refocus on studying much simpler Hamiltonians}''. 
We demonstrate that an RVB liquid phase, exhibiting spin-charge separation, emerges naturally upon dilute-doping a large-$U$ (on-site Coulomb repulsion) Hubbard model on a lattice of corner-sharing tetrahedra -- be it the planar checkerboard or the three-dimensional pyrochlore lattice -- near half-filling. This result hinges on kinematic spin correlations set by frustrated hopping and is an instance of the so-called `counter-Nagaoka effect'~\cite{Haerter2005}. 
Without loss of generality, in the following we shall consider the case of hole doping; the same can be said for electron doping when the hopping amplitude has the opposite sign. 
We prove our claims by finding an exact lower bound to the Hamiltonian energy, and by explicitly constructing an RVB liquid state that precisely matches it in the thermodynamic single-hole limit. 
We confirm this behaviour by investigating numerically systems of finite size.
(A numerical study of one- and two-hole doping in the $t$-$J$ model on the checkerboard lattice had already been carried out by Poilblanc~\cite{Poilblanc2004}, where $t$ is the electron hopping amplitude and $J$ the nearest-neighbour exchange interaction strength. Poilblanc's work highlighted the $-4t$ and $-8t$ respective ground state energies, and arguing how the singlet background somehow changes the effective hopping -- a precursor of the understanding in terms of RVB liquid and two $\pi$-flux attachment presented in this work.)
Here we propose a way to construct single-hole and two-holes RVB liquid ground state wave functions for any system size. These seem to be exact within numerical accuracy, a truly notable feat for a strongly correlated electron problem. Finally, we show numerically that our results are stable in systems of finite hole density, as well as in presence of sufficiently small exchange interactions between the spins.

The crucial role of kinetic energy frustration in stabilising a topologically ordered RVB liquid state is a remarkable outcome of our findings, which provide a hitherto-unexplored way to connect dimer coverings with solutions to the Hubbard model. Our results have the potential to shift the traditional paradigm, which has predominantly considered exchange frustration as the mechanism for stabilising quantum spin liquids. In other words, doping insulators that lie deep within the Mott regime with \emph{a single type of carrier} (electrons or holes, depending on the sign of the hopping amplitude) promises to become a guiding principle for the discovery of spin liquid states in correlated matter. 

Our Hamiltonian is eminently simple and realistic, and of direct relevance to several experimental settings as well as synthetic quantum platforms~\cite{Feynman1982,Trabesinger2012,Georgescu2014}. We identify in particular some families of pyrochlore compounds as suitable frameworks to experimentally test our predictions and suggest how it may be possible to manifest the elusive RVB state 
%for the first time 
in these materials.
%
%
%%%%%%%%%%%%%%%%%%%%%%%%%%%%%%%%%%%%%%%%%%%%%%%%%%%%%%%%%

\section{Results
\label{sec:results}}
We study the $t$-$J$ model on a lattice of corner-sharing tetrahedra 
%~\cite{footnote:checkerboard} 
%~\footnote{Our results apply straightforwardly, e.g., to the 2D case of the checkerboard lattice, or to the 3D case of the pyrochlore lattice. Note that for the 16-site system considered below, the two lattices in fact coincide.} 
(see Fig.~\ref{fig:pyrochlore lattice}) near half-filling, with Hamiltonian
\begin{equation}
    \hat{\mathcal{H}} 
    = 
    - t\sum_{\langle i,j\rangle \; \sigma} \left[
      \hat{c}^\dagger_{i\sigma} \hat{c}_{j\sigma} + h.c. 
    \right] 
    + J\sum_{\langle i, j \rangle} \hat{\bm S}_i\cdot \hat{\bm S}_j
\, ,
\label{eq:ham}
\end{equation}
where $\langle i,j\rangle$ denotes nearest-neighbour pairs of sites; $\hat{c}^\dagger_i$ ($\hat{c}_i$) is the constrained fermionic creation (annihilation) operator at site $i$, acting in the space where double occupancy is strictly forbidden (see Supp.~Mat.~Note~\ref{app:methods}); and $\hat{\bm S}_i$ is the corresponding spin operator, $\hat{\bm S}_i = \tfrac{1}{2}\sum_{\sigma, \sigma^\prime} \hat{c}^\dagger_{i\sigma} {\bm \sigma}_{\sigma\sigma'} \hat{c}_{i\sigma'}$. 
Here $\sigma$ labels the states of spin-$1/2$ degrees of freedom, and ${\bm \sigma}$ is the conventional vector of Pauli matrices. 
For generality, we treat the hopping $t$ and the spin exchange $J$ as independent parameters.
Note that for fermions and our convention in Eq.~\eqref{eq:ham}, hole doping with $t>0$ or electron doping with $t < 0$ correspond to frustrated hopping when the elementary loop of the lattice has an odd number of sites. In the following we shall focus, without loss of generality, on the case of hole doping and $t>0$. 

Our results apply straightforwardly, e.g., to the 2D case of the checkerboard lattice, or to the three-dimensional case of the pyrochlore lattice. Note that for the 16-site system considered below, the two lattices in fact coincide. 

Our first result is analytical, derived in the thermodynamic limit for a single hole and $J=0$. As demonstrated in detail in the Supp.~Mat.~Note~\ref{app:proofs}, the spectrum of the Hamiltonian in Eq.~\eqref{eq:ham} has a strict lower bound of $-4t$. Furthermore, we prove that this bound is met by the RVB spin liquid ground state
\begin{equation}
| \Psi_{\rm RVB} \rangle 
\propto 
\sum_{{\bm r},\{  a_{\bm r} \} } \prod_{\langle i,j \rangle \in a_{\bm r}} \hat{d}^\dagger_{ij} | 0 \rangle
\, , 
\label{eq:GSwf}
\end{equation}
where ${\bm r}$ spans all possible positions of the holon; $a_{\bm r}$ is a generic dimer covering of the lattice with a holon at ${\bm r}$ subject to the condition that there is one and only one dimer per tetrahedron (see e.g., Fig.~\ref{fig:pyrochlore lattice}); and 
$\hat{d}^\dagger_{ij} = \frac{1}{\sqrt{2}} (\hat{c}^{\dagger}_{i \uparrow} \hat{c}^{\dagger}_{j \downarrow} - \hat{c}^{\dagger}_{i \downarrow} \hat{c}^{\dagger}_{j \uparrow} )$. 
We shall refer to each state in the superposition forming $| \Psi_{\rm RVB} \rangle$ as a dimer-singlet state. 
The kinetic energy frustration disappears for the opposite sign of the hopping amplitude, which is equivalent to inserting a $\pi$-flux on each triangular plaquette of the pyrochlore lattice. The RVB wave function minimises the holon kinetic energy by spontaneously generating a $\pi$-flux on the triangular face containing the hole and a singlet. Namely, the RVB wave function binds two $\pi$-fluxes (fluctuating with the dimer-singlet resonant states) to each hole to lower its kinetic energy to $-4t$ by partially lifting the frustration (see Fig.~\ref{fig:pyrochlore lattice})~\cite{Poilblanc2004}. 

This is an instance of the so-called counter-Nagaoka effect~\cite{Haerter2005}, whereby frustrated quantum hole hopping leads to antiferromagnetic correlations in a Mott insulator near half-filling. The phenomenon has been shown to generically lead to conventionally ordered phases in two dimensions~\cite{Sposetti2014,Lisandrini2017,Zhang2018,Morera2024}, 
with the notable exception of a Husimi cactus of triangles (akin locally to a kagome lattice) where Kim proved that it leads to a regular valence bond pattern in presence of a single (deconfined) hole~\cite{Kim2023}. 
It had not been thus far investigated in three dimensions, and it is all the more remarkable to see it favour a quantum spin liquid state in lattices of corner-sharing tetrahedra (for a brief review on the counter-Nagaoka effect, see Supp.~Mat.~Note~\ref{app:counterNagaoka}). 

Note that, unlike the triangular case considered in Ref.~\onlinecite{Kim2023}, where the dimer-singlet configuration is fully determined once the hole position is fixed, for our corner-sharing tetrahedra there are extensively many dimer-singlets forming a massively entangled superposition of electronic states for each fixed hole position. The latter is exactly what one expects of an RVB liquid state.

\begin{figure}[ht!]
\begin{center}
\includegraphics[width=\columnwidth]{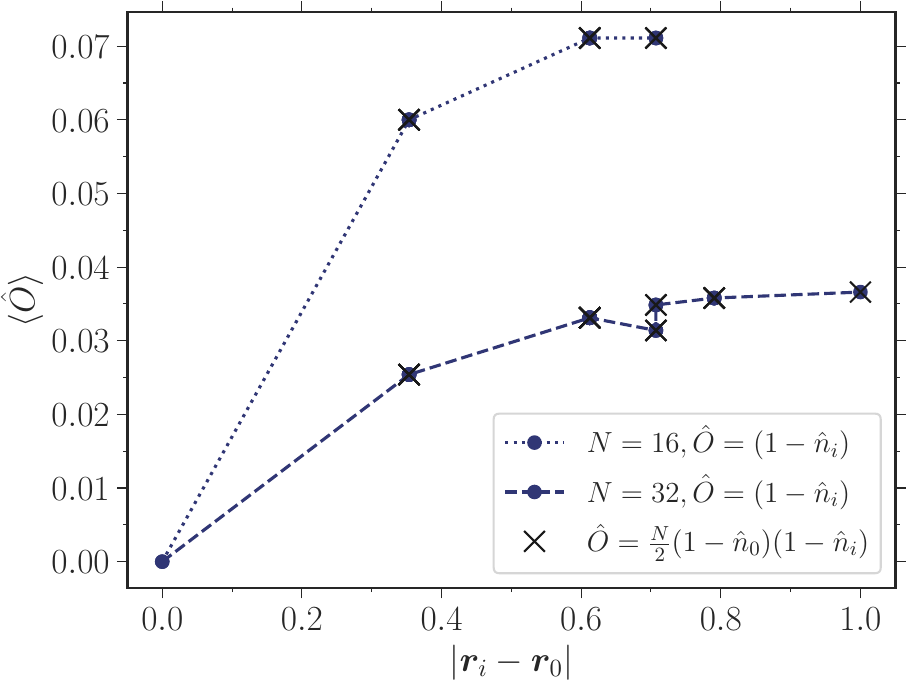}
\\ \hspace{1.3cm}
\includegraphics[width=\columnwidth]{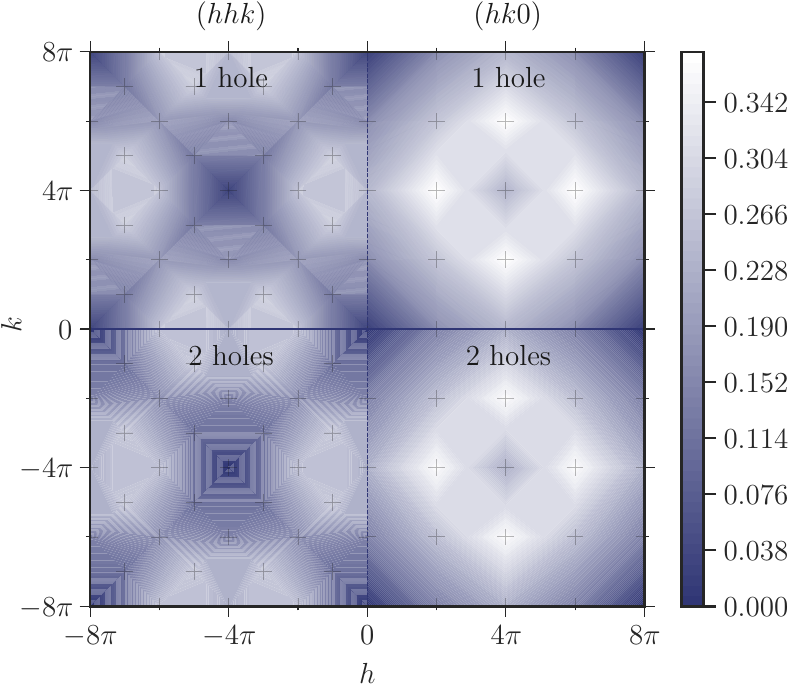}
\end{center}
\caption{{\it Correlations and spin structure factor in the RVB state}. Top, blue dots show the holon density $\langle \hat{O}\rangle = \langle 1-\hat{n}_i\rangle$ ($\hat{n}_i$ being the electron density operator) for a 16-site and a 32-site pyrochlore system doped with a single hole as a function of distance from the pinned spin,
 $\vert \bm{r}_i - \bm{r}_0 \vert$, for system sizes $N = 16, \, 32$.
It indicates that the spinon and holon separate (the two different values at distance $\sim 0.7$ are due to the two inequivalent third-nearest-neighbour sites on the pyrochlore lattice). Crosses show the holon-holon correlations when there are two holes present in the system. They are numerically identical to the holon-spinon correlations.
Bottom, spin structure factor in the $(hhk)$ and $(hk0)$ reciprocal lattice planes for the 32-site pyrochlore system with one hole (top half) and two holes (bottom half), for $t = 1$ and $J=0$. 
Accessible momenta are marked by grey pluses.}
\label{fig:one_hole_corr}
\end{figure}

The ground state Eq.~\eqref{eq:GSwf} is exact in the thermodynamic limit, where such dimer coverings with a single monomer are well-defined. In a finite lattice of corner-sharing tetrahedra with periodic boundary conditions, introducing a hole necessarily requires an unpaired spin. As expected for an RVB ground state, the spin and the charge of the fermion added to the Mott insulator separate, i.e., the probability of finding the holon within a finite distance from the spinon is zero in the thermodynamic limit. The spinon must necessarily occupy a passive site of the tetrahedron, implying that the holon will never visit the site occupied by the unpaired spin. 
In other words, the spinon is expected to have infinite mass (static spinon) and the spinon-holon interaction is expected to be repulsive. Both observations are supported by our numerics, which give our second set of results and attest the stability of our RVB liquid phase in finite-size systems, finite hole density, and the presence of interactions. 

We studied the behaviour of Eq.~\eqref{eq:ham} in finite pyrochlore systems using exact diagonalisation (ED) on a 16-site cubic cell, and density matrix renormalisation group (DMRG)~\cite{White1992,Schollwock2005,Schollwock2011} on a 32-site system, with periodic boundary conditions (for details of the study, see Supp.~Mat.~Note~\ref{app:correlations}). 
Remarkably, we find that the ground state energy bound $-4t$ is met (when $J=0$) for all system sizes, within numerical precision. A non-trivial degeneracy is observed ($18$ states for 16 sites and $34$ states for 32 sites). Consistently with the analytical considerations above, we find a unique ground state (again with energy $-4t$) when the unpaired spin is pinned to a given lattice site. 

Within numerical accuracy, we show that the ground state wave function with a fixed spin is given by the equal amplitude superposition of all dimer-singlet configurations (constrained to no more than one dimer per tetrahedron) with a delocalised holon, for our 16-site and 32-site systems (see Supp.~Mat.~Note~\ref{app:correlations}). This allows us to propose an exact wave function for the single-hole case for all system sizes -- which is a highly unusual feat. 
Such a state has vanishing expectation value of the projector onto the maximum angular momentum for each tetrahedron with no hole~\cite{Batista2004,Raman2005} (see also Kivelson, S.A., private communication, unpublished), which we verified to be the case (Supp.~Mat.~Note~\ref{app:correlations}). 

As illustrated in the top panel of Fig.~\ref{fig:one_hole_corr}, the holon density shows weak repulsive dependence on the distance from the spinon in finite size systems, in agreement with spin-charge separation in the thermodynamic limit, argued earlier on analytical grounds (see also Supp.~Mat.~Note~\ref{app:correlations}). 
In the bottom panel of Fig.~\ref{fig:one_hole_corr}, we show the spin structure factor of our numerical ground states. Characteristic differences that could be measured experimentally seemingly distinguish it from, for example, the classical or quantum Heisenberg spin liquid structure factors (see Supp.~Mat.~Note~\ref{app:correlations}); however future work on larger system sizes is needed to make more precise statements about this. 
On general grounds, we expect exponentially decaying spin correlations; the nature of the singlet correlations is less clear, as the correspondence with spin ice states~\cite{Baskaran1988,Read1989,Fradkin1990,Raman2005} could induce power law behaviour and an emergent U(1) gauge symmetry with gapless photon excitations; this is however far from trivial due to the non-orthogonality of the dimer-singlet states. 

We briefly studied the case $J=0$ and two holes, which we were able to access numerically in the 16-site and 32-site systems. We observe no tendency towards holon attraction (see top panel of Fig.~\ref{fig:one_hole_corr} and Supp.~Mat.~Note~\ref{app:twoholes}). The ground state is now unique, with energy $-8t$, suggesting that the second holon is able to occupy the site of the unpaired spin, and to render it itinerant irrespective of the presence of the other holon. Consistently, we find that the holon-spinon correlations for a single hole are numerically identical to the holon-holon correlations for two holes (Fig.~\ref{fig:one_hole_corr}). 
The nature of the two-holes ground state can be understood from the perspective of the $\pi$-flux binding discussed earlier. A holon can only move on a tetrahedron around the triangular face that contains a singlet (see Fig.~\ref{fig:pyrochlore lattice}), and it is forbidden by interference effects to visit the fourth tetrahedral site (thus rendering two holons on the same tetrahedron blind to one another). In other words, the single-hole RVB wave function is an equal weight superposition of the holon propagating in \emph{all possible sublattices} that can be constructed by choosing one triangle of each tetrahedron that is visited. For the two-holes RVB wave function (namely, the equal amplitude superposition of all dimer-singlet states hosting two holons, with no more than one dimer per tetrahedron), the holons propagate through different triangles when they meet at the same tetrahedron, meaning that the effective coordination number is still equal to 4 for each of them, and the energy is $2 \times (-4t) = -8t$. 
These observations, in particular the coincidence of the spinon-holon and holon-holon correlators in Fig.~\ref{fig:one_hole_corr}, are strongly suggestive that the holon excitations in our RVB state are bosonic.

We verified numerically that the proposed RVB wave function is also the exact ground state for the two-holes system (for 16 sites and 32 sites). 
The argument does not extend to three or more holes on finite systems, and their behaviour is left to future investigations. 

Intriguingly, the opposite (stoichiometric) limit of vanishing hole density is singular for the 16-site and 32-site systems in question. The wave function given by the equal amplitude superposition of all dimer-singlet states with one and only one dimer per tetrahedron vanishes identically 
%~\cite{footnote:vanishing}
%~\footnote{We note that the vanishing occurs also for equal amplitude superpositions of all dimer-singlet states, without the constraint of one dimer per tetrahedron --- a curious fact that, to the best of our knowledge, had not been pointed out in the literature to date.} 
(see Supp.~Mat.~Note~\ref{app:proofs}). 
We note that the vanishing occurs also for equal amplitude superpositions of all dimer-singlet states, without the constraint of one dimer per tetrahedron --- a curious fact that, to the best of our knowledge, had not been pointed out in the literature to date.
The RVB liquid phase therefore appears to exist solely in presence of hole doping (at least within the remit of our accessible system sizes and periodic boundary conditions), and it can only be stabilised by kinetic energy frustration. 

\begin{figure}
\begin{center}
\includegraphics[width=\columnwidth]{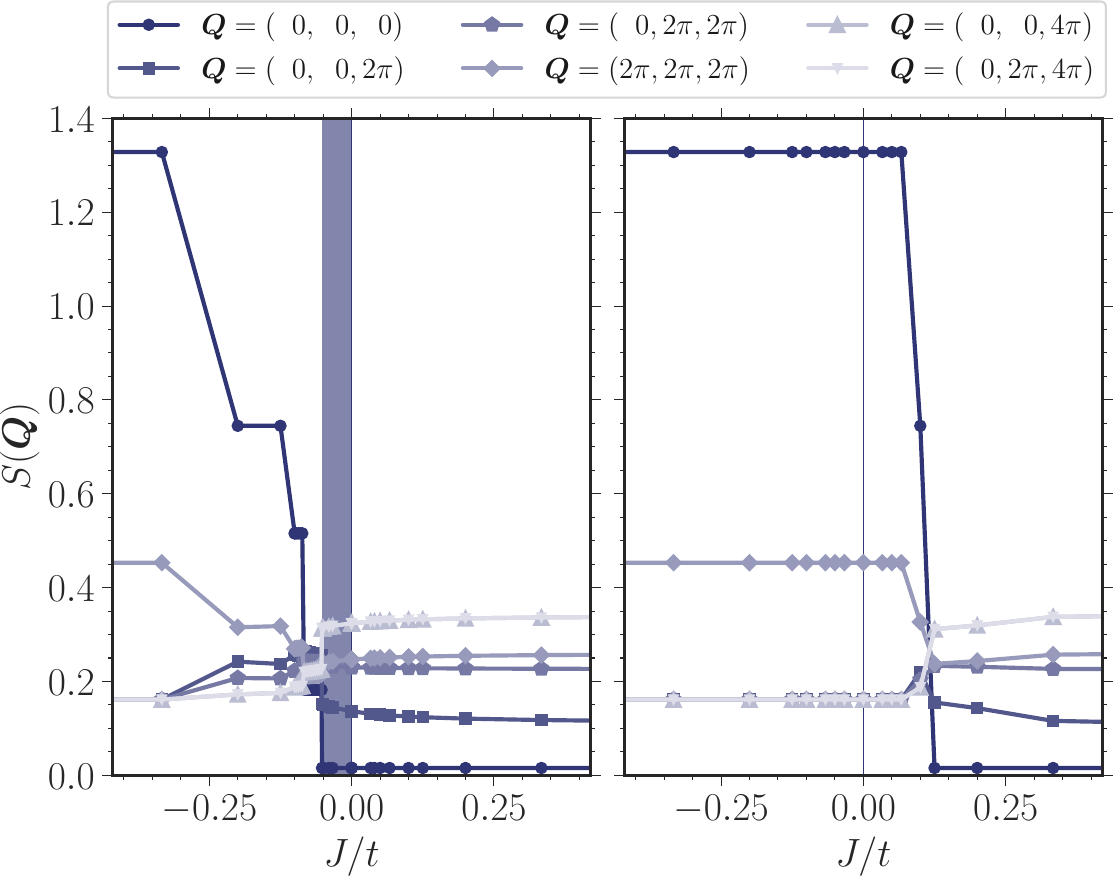}
%hole_vs_electron.pdf}
\end{center}
\caption{{\it Spin structure factor}. Dependence of the ground state spin structure factor $S(\bm{Q})$ (for all individual reciprocal lattice points $\bm{Q}$) on $J/t$ ($t > 0$) for a 16-site system. Left, the system is doped with one hole.
The region where the correlations are antiferromagnetic, despite the ferromagnetic $J$, is shaded. 
Right, the system is doped with one electron, for comparison.}
\label{fig:J/t_one_hole}
\end{figure}

Finally, we turn to look at the concomitant effects of $t$ and $J$ in the system. We consider the case of a 16-site system with a single hole; we fix $t=1$ and vary $J$, spanning both positive and negative values. 
In Fig.~\ref{fig:J/t_one_hole}, we plot the behaviour of the ground state spin structure factor for all distinct points in reciprocal space. 
We find that the spin liquid behaviour persists down to a finite negative value of $(J/t)_c \simeq -0.052$ (for holon density $6.25$\%), suggesting that a transition from ferromagnet to quantum spin liquid (QSL) may be induced on the pyrochlore lattice by frustrated hole doping.
The effective doping-induced antiferromagnetic interaction between the spins has been argued to be proportional to the holon density at sufficiently small concentrations~\cite{Haerter2005}, and we expect our results to correspondingly become more pronounced (with the caveat of possible phase separation effects as in Ref.~\onlinecite{Eisenberg2002} for the large but finite $U$ Hubbard limit). 
Indeed, adding a second hole to the 16-site system, we find $(J/t)_c \simeq - 0.173$ (holon density $12.5$\%), see Supp.~Mat.~Note~\ref{app:twoholes}. 
In Fig.~\ref{fig:J/t_one_hole}, we further compare for reference the case of electron doping, which is not frustrated for $t > 0$ and exhibits a kinetic tendency towards ferromagnetic order through the Nagaoka effect. 

%
%
%%%%%%%%%%%%%%%%%%%%%%%%%%%%%%%%%%%%%%%%%%%%%%%%%%%%%%%

\section{Discussion}
\label{sec:discussion}
Our results provide a simple and elegant solution to a problem that has eluded researchers for nearly half a century. Through exact results and detailed numerical simulations, we demonstrate that the long-sought-after RVB spin liquid state exhibiting spin-charge separation emerges naturally as ground state of a large-$U$ Hubbard Hamiltonian on lattices of corner-sharing tetrahedra, upon dilute-doping away from the Mott insulating state. The key aspect of this emergent behaviour is the kinetic energy frustration of carriers that result from doping the Mott state. 

The non-Fermi-liquid behaviour (e.g., spin-charge separation) of such an RVB liquid state implies that the large-$U$ limit of the pyrochlore Hubbard model cannot be smoothly connected to its weak-coupling counterpart (small $U$). Hence, an effective description at large $U$, which is non-perturbative in nature, is essential to fully characterise it. It is interesting to contrast it to the Fermi liquid behaviour argued to occur in spin liquid states stabilised by frustration in lattices of corner-sharing tetrahedra near half-filling in finite-$U$ Hubbard models (see e.g., Refs.~\onlinecite{Fujimoto2001,Pan2023}). This raises the intriguing possibility of a crossover between the two regimes as a function of doping. 

The eminently neat and realistic nature of the Hamiltonian opens the door to further theoretical investigations into the physics of RVB phases, possibly also with the help of implementations on synthetic quantum platforms~\cite{Trabesinger2012,Georgescu2014} (note in particular how the leading RVB liquid behaviour is already apparent in relatively small systems of 16 sites). 
The rare availability of an exact single-hole and two-hole ground state wave function for systems of any size provides uncommon ability to compare theory and simulator experiments. 

It should also be possible to test our results in real materials. In the first instance, an experimental comparison between the effects of hole and electron doping, i.e., switching between counter-Nagaoka antiferromagnetic correlations and Nagaoka ferromagnetic ones~\cite{Ciociaro2023} (as shown in Fig.~\ref{fig:J/t_one_hole}), could be used to check if a given Mott insulator is indeed in the large-$U$ regime relevant to this work. 
The dynamical spin structure factor $S({\bm Q}, \omega)$ of the predicted RVB state can be experimentally accessed via neutron scattering, as a function of reciprocal lattice vector $\bm{Q}$ and angular frequency $\omega$. The low-energy region, in particular, ought to include a spinon continuum, flat in the infinite-$U$ limit and broadened by $J$, which could be used to measure spin interactions. The spin-spin correlation length can be directly extracted from the static spin structure factor $S({\bm Q}) = \int d\omega S({\bm Q}, \omega)$.
Spin-charge separation can be indirectly measured via photo-emission experiments, which will reveal a vanishing weight of the quasi-particle peak and can also be used to extract the hopping amplitude $t$.
Finally, in presence of interactions (e.g., exchange $J$) the undoped Mott insulating phase is likely to exhibit magnetic properties that are notably different from those of the RVB liquid phase. A transition between the two phases as a function of doping could be detected via signatures in thermodynamic properties, such as specific heat and susceptibility, as well as in neutron scattering or muon-spin rotation measurements, raising the tantalising prospect of observing hole-induced quantum spin liquid behaviour. 
If the interactions are ferromagnetic, one could envisage looking for polaronic behaviour and possibly magnonic Cooper pairs (bound states of two fermions and one magnon) in the ferromagnetic phase close to the transition to the RVB state~\cite{Zhang2018,nazaryan2024}. 

Candidate materials that realise large-$U$-Hubbard-model Mott insulators on lattices of corner-sharing tetrahedra are not commonplace. One can nonetheless identify families of compounds which, while they do not immediately map onto our model, provide promising avenues for further investigation and the possible realisation of the RVB liquid behaviour presented in our work. 
One such family of compounds, the $5d$ transition metal pyrochlore oxides, has been extensively studied and is known to form regular pyrochlore lattices. Rare-earth iridates and ruthenates, RE$_2$TM$_2$O$_7$, host $5d^5$ TM$^{4+}$ ions with an effective angular momentum $J_{\rm eff} = 1/2$ ground-state doublet. Holons can be introduced by doping on the non-magnetic RE site, thus effectively replacing TM$^{4+}$ with TM$^{5+}$, $J_{\rm eff} = 0$ ions. Some doping experiments of these materials have already been performed, including measurements on polycrystalline Y$_{2-x}$Ca$_x$Ir$_2$O$_7$ for $x = 0$--$0.2$~\cite{Fukazawa2002,Zhu2014} where a marked effect on the magnetic and transport properties was observed. However, the polycrystalline nature of the samples used in these studies seems to have led to confusing and contradictory results (see Supp.~Mat.~Note~\ref{app:expm} for additional details). 
%cite{supmat}. 
Recent progress in producing high-quality single crystals of this material family (see, e.g., Refs.~\onlinecite{Ishikawa2012,Cathelin2020,Uehara2022} and inset to Fig.~\ref{fig:pyrochlore lattice}), could provide the opportunity to understand these changes and perhaps observe evidence for the predictions discussed above. (See Supp.~Mat.~Note~\ref{app:expm} for a more detailed review of the materials that could potentially be investigated, the methods by which they could be grown, and possible experimental routes to measure a doping-induced order-to-QSL transition.) Having said this, certain caveats should be flagged regarding the applicability of our theoretical results to these $5d$ compounds: they tend to be weak insulators, and they display strong spin-orbit effects; hence they are less likely to be well-described by models of isotropic hopping. Further theoretical work is needed to study the tolerance of the original model to these departures.
Another family of compounds involves $3d$ transition metals, which offer large $U/t$ ratios and (owing to weak spin-orbit coupling) near-isotropic hopping intrinsic to our model. 
Investigating pyrochlore lattices of spin-1/2 Cu(II) is particularly promising for two reasons: they are strong Mott insulators with a large $U/t$ ratio, and their spin-orbit coupling is substantially weaker than that of 4$d$ and 5$d$ elements. However, only a few suitable systems have been identified to date (see e.g., Refs.~\onlinecite{Pinsard2004,Wills2004,Uematsu2007,Kawabata2007,Nishiyama2011}) and these often exhibit distortions from the ideal pyrochlore lattice. The impact of such distortions on our model would need to be carefully considered. 

Our results apply to corner-sharing tetrahedral lattices in general, including the two-dimensional checkerboard lattice~\cite{Poilblanc2004} (see Supp. Mat. Note~\ref{app:checkerboard} for a study of 16- to 48-site checkerboard systems). While less common than the pyrochlore lattice in real materials, one wonders whether such geometry could emerge in quasi-2D square lattice compounds, possibly mediated by inter-layer atoms; or perhaps in 2D materials and synthetic quantum platforms. From a theoretical perspective, it would be interesting to investigate in greater detail the robustness of the RVB liquid behaviour and spin-charge separation to an asymmetry between the nearest-neighbour and next-nearest-neighbour (diagonal) hopping terms on the checkerboard lattice. It would also be interesting to study the effects of non-magnetic vacancy disorder, which has been recently argued to destabilise some but not all short-range RVB liquid phases in 2D, depending on the lattice~\cite{Ansari2024}. 

As speculated earlier, the RVB singlet state in our system could exhibit an emergent $U(1)$ gauge symmetry with power-law singlet correlators and gapless (photon) excitations (in three dimensions)~\cite{Baskaran1988,Read1989,Fradkin1990}. 
In presence of itinerant (fermionic~\cite{Read1989b}) holons, this is reminiscent of models used to investigate Fermi Liquid Star (FL*) behaviour~\cite{Senthil2003}, and may provide an additional experimental avenue to explore it. 

In closing, our work proposes kinetic energy frustration of doped Mott insulators as a route to realising robust topologically ordered states, in contrast to the conventional approach based on geometrically frustrated interactions. This holds great promise for the theoretical and experimental discovery of emergent states of matter. As for a potential connection between kinetically frustrated RVB liquids and superconductivity~\cite{Anderson1987}, this shall remain an open question for future work. 
%
%
%%%%%%%%%%%%%%%%%%%%%%%%%%%%%%%%%%%%%%%%%%%%%%%%%%%%%%%

\section*{Acknowledgements} 
We would like to thank J.Chalker, V.Oganesyan, S.Powell, and O.F.Sylju{\aa}sen for useful discussions. 
C.G. acknowledges support from the Aker Scholarship. 
The work of A.\v{S}. is supported by the European Union’s Horizon Europe research and innovation programme under the Marie Sk\l{}odowska-Curie Actions Grant agreement No. 101104378. 
D.P. would like to acknowledge the Oxford-ShanghaiTech Collaboration Project for financial support. 
C.D.B. acknowledges support from the U.S.~Department of Energy, Office of Science, Office of Basic Energy Sciences, under Award Number DE-SC0022311. 
This work was supported in part by the Engineering and Physical Sciences Research Council (EPSRC) grants No.~EP/T028580/1 and No.~EP/V062654/1 (C.C.). 
For the purpose of open access, the authors have applied a Creative Commons Attribution (CC-BY) licence to any
Author Accepted Manuscript version arising from this
submission.
%
%
%%%%%%%%%%%%%%%%%%%%%%%%%%%%%%%%%%%%%%%%%%%%%%%%%%%%%%%

\section*{Author contributions}
The project was conceived by C.C. and C.G., and all authors contributed to its development, with specific focus by C.G. on the numerical efforts with A.\v{S}. as discussion partner, by C.D.B. on the analytical proofs, and by P.A.G. and D.P. on the experimental component. All authors contributed to writing the manuscript. 
%
%
%%%%%%%%%%%%%%%%%%%%%%%%%%%%%%%%%%%%%%%%%%%%%%%%%%%%%%%

\section*{Competing interests}
The authors declare to have no competing interests in this work.
%
%
%%%%%%%%%%%%%%%%%%%%%%%%%%%%%%%%%%%%%%%%%%%%%%%%%%%%%%%

\section*{Materials and Correspondence}
Correspondence should be addressed to C.C., and queries about the numerical results should be addressed to C.G. 
%
%
%%%%%%%%%%%%%%%%%%%%%%%%%%%%%%%%%%%%%%%%%%%%%%%%%%%%%%

\putbib
\end{bibunit}
%
%
%%%%%%%%%%%%%%%%%%%%%%%%%%%%%%%%%%%%%%%%%%%%%%%%%%%%

\begin{bibunit}

\section{Methods
\label{sec:methods}}
We studied a 16-site system consisting of a single conventional cubic unit cell of the pyrochlore lattice (see Supp.~Mat.~Note~\ref{app:methods} for illustrations) using ED with the Lanczos method to compute the lowest eigenvalues and corresponding eigenstates in each magnetisation sector. 
We also considered a 32-site pyrochlore system consisting of a $2 \times 2 \times 2$ cell constructed from the fcc primitive lattice vectors. For this system size, we used two-site DMRG with an energy convergence criterion of $10^{-10}$ and a maximum bond dimension of $12800$. We imposed periodic boundary conditions and created a 1D snake path of the system following the numbering shown in the Supp.~Mat.~Note~\ref{app:methods}. The choice of snake path seems to have little impact on the convergence for our highly entangled 3D system.
The DMRG calculations were performed using the TeNPy Library (version 0.10.0) [J. Hauschild and F. Pollmann, SciPost Phys. Lect. Notes
, 5 (2018), code available from https://github.com/
tenpy/tenpy].
%~\cite{tenpy}. 

A single DMRG calculation gives one ground state. When the ground state is degenerate, we biased each ground state and compute correlations as an average over all ground states. 
In the case of pure hopping ($J=0$), the full ground state manifold was found by pinning the unpaired spin by adding an on-site magnetic field. This gives an overcomplete set of ground states, and a linearly independent set of ground states is then found via Gram-Schmidt decomposition. 

A similar approach was deployed to study our system on the checkerboard lattice, using ED for 16 sites (1 and 2 holes) and 24 sites (1 hole), and DMRG for 24 sites (2 holes), 36 sites (1 and 2 holes) and 48 sites (1 and 2 holes). 
%
%
%%%%%%%%%%%%%%%%%%%%%%%%%%%%%%%%%%%%%%%%%%%%%%%%%%%%

\section{Data availability}
All data shown in the figures can be derived using computer simulations with parameters given in the manuscript and Supplementary Material. 
%
%
%%%%%%%%%%%%%%%%%%%%%%%%%%%%%%%%%%%%%%%%%%%%%%%%%%%%

\section{Code availability}
All code used in this work is readily available through libraries as discussed in the Methods section. 
%
%
%%%%%%%%%%%%%%%%%%%%%%%%%%%%%%%%%%%%%%%%%%%%%%%%%%%%%%%

%\putbib
\end{bibunit}
%
%
%%%%%%%%%%%%%%%%%%%%%%%%%%%%%%%%%%%%%%%%%%%%%%%%%%%%%%%

\clearpage
\newpage
%
%
%%%%%%%%%%%%%%%%%%%%%%%%%%%%%%%%%%%%%%%%%%%%%%%%%%%%%%
\appendix
%%%%%%%%%%%%%%%%%%%%%%%%%%%%%%%%%%%%%%%%%%%%%%%%%%%%%%

\setcounter{figure}{0}
\renewcommand{\thefigure}{S\arabic{figure}}
\renewcommand{\appendixname}{SM Note}

\begin{bibunit}

\iffalse
%\begin{widetext}
\onecolumngrid
\vspace{\columnsep}
\begin{center}
{\Large\bf Supplementary Material for:} 

\vspace{0.2 cm}
{\large\bf ``Spin-charge separation and resonant valence bond spin liquid in a frustrated doped Mott insulator''}

\vspace{0.5 cm}
Cecilie Glittum,$^{1}$
Antonio \v{S}trkalj,$^{1,2}$ 
Dharmalingam Prabhakaran,$^{3}$ 
Paul A. Goddard,$^{4}$ 
Cristian D. Batista,$^{5,6}$ 
and 
Claudio Castelnovo$^{1}$

\vspace{0.3 cm}
$^{1}$T.C.M. Group, Cavendish Laboratory, JJ Thomson Avenue, Cambridge CB3 0HE, United Kingdom

$^{2}$Department of Physics, Faculty of Science, University of Zagreb, Bijeni\v{c}ka c. 32, 10000 Zagreb, Croatia

$^{3}$Department of Physics, University of Oxford, Clarendon Laboratory, Parks Road, Oxford OX1 3PU, UK

$^{4}$Department of Physics, University of Warwick, Coventry CV4 7AL, United Kingdom

$^{5}$Department of Physics and Astronomy, University of Tennessee, Knoxville, TN 37996, USA

$^{6}$Neutron Scattering Division, Oak Ridge National Laboratory, Oak Ridge, TN 37831, USA
\end{center}
\vspace{\columnsep}
\twocolumngrid
%\end{widetext}
\fi
\onecolumngrid
\vspace{\columnsep}
\begin{center}
{\Large\bf Supplementary Material Table of Content} 

\vspace{0.6 cm}
\begin{tabular}{ll}
Supp.~Mat.~Note~\ref{app:proofs}: & Proofs
\\
Supp.~Mat.~Note~\ref{app:counterNagaoka}: & The counter-Nagaoka effect
\\
Supp.~Mat.~Note~\ref{app:methods}: & Models and methods
\\
Supp.~Mat.~Note~\ref{app:correlations}: & Doping and correlations
\\
Supp.~Mat.~Note~\ref{app:twoholes}: & Two holes
\\
Supp.~Mat.~Note~\ref{app:checkerboard}: & Checkerboard
\\
Supp.~Mat.~Note~\ref{app:expm}: & Experiments
\end{tabular}
\end{center}
\vspace{\columnsep}
\twocolumngrid
%
%
%%%%%%%%%%%%%%%%%%%%%%%%%%%%%%%%%%%%%%%%%%%%%%%%%%%%%%%

\section{Proofs
\label{app:proofs}}
In this section, we present the proof of the first result in our work, relating to the ground state (GS) of Eq.~\eqref{eq:ham} in the main text, for a single hole in the thermodynamic limit when $J=0$ and $t>0$ (frustrated hopping).
We note that the proof applies generally to lattices of corner-sharing tetrahedra, including the three-dimensional pyrochlore lattice, as well as the two-dimensional checkerboard lattice. 
%
%
%-----------------------------------------------------

\subsection{Ground state energy
\label{app:proofsE0}}
Since the Hamiltonian $\hat{\mathcal{H}}$ includes only nearest neighbour hopping terms, it can be decomposed as a sum of Hamiltonians $\hat{\mathcal{H}}^{\alpha}$ acting on each tetrahedron $\alpha$:
\begin{equation}
\hat{\mathcal{H}} = \sum_{\alpha} \hat{\mathcal{H}}^{\alpha}
\, ,
\end{equation}
where the dual lattice index $\alpha$ labels the different tetrahedra and $\hat{\mathcal{H}}^{\alpha}$ includes only the six bonds of the tetrahedron $\alpha$. 

The ground state $| \Psi_0 \rangle$ can be expressed as 
\begin{equation}
| \Psi_0 \rangle = \sum_{ {\bm r}} | \phi_{\bm r} \rangle
\, ,
\end{equation}
where $|\phi_{\bm r} \rangle$ represents a state where the holon occupies the site ${\bm r}$ and the rest of the spins are in an arbitrary spin configuration,
$
\langle \phi_{\bm r} | \phi_{{\bm r}'} \rangle = \delta_{{\bm r},{\bm r}'}    \langle \phi_{\bm r} | \phi_{{\bm r}} \rangle
$,
and $\langle \Psi_0  | \Psi_0  \rangle=1 $.

After introducing  the states
\begin{equation}
| \chi_{\alpha}  \rangle  = \sum_{{\bm r} \in \alpha} | \phi_{\bm r} \rangle
\, ,
\end{equation}
where the sum runs over the $4$ sites of the tetrahedron $\alpha$, we compute the expectation value of the energy,
\begin{equation}
E_g = \sum_{  {\bm r}, {\bm r}'}  \langle \phi_{\bm r} | \hat{\mathcal{H} } |\phi_{\bm r'} \rangle = 
\sum_{  \langle {\bm r}, {\bm r}'\rangle }  \langle \phi_{\bm r} | \hat{\mathcal{H} } |\phi_{\bm r'} \rangle =
\sum_{\alpha}     \langle \chi_{\alpha} | \hat{\mathcal{H} }^{\alpha} | \chi_{\alpha} \rangle
\, ,
\label{eq:exp}
\end{equation}
where $\langle {\bm r}, {\bm r}'\rangle$ restricts the sum to nearest-neighbour sites.
By introducing the probability distribution over tetrahedra:
\begin{equation}
p_{\alpha} = \frac{1}{2}  \langle   \chi_{\alpha} | \chi_{\alpha} \rangle
\, ,
\end{equation}
where the normalisation condition $\sum_{\alpha} p_{\alpha} =1$ follows from $\langle \Psi_0 | \Psi_0  \rangle=1$, 
we express $E_g$ as
\begin{equation}
E_g = \sum_{\alpha}  p_{\alpha}  2 \frac{\langle \chi_{\alpha} | \hat{\mathcal{H} } | \chi_{\alpha} \rangle}{\langle   \chi_{\alpha} | \chi_{\alpha} \rangle}
\, .
\end{equation}

Now, assume that $E_g< -4t$. Since $p_\alpha$ is a probability distribution over tetrahedra, this assumption implies that there is at least one tetrahedron $\beta$ for which
\begin{equation}
2 \frac{\langle \chi_{\beta} | \hat{\mathcal{H} } | \chi_{\beta} \rangle}{\langle   \chi_{\beta} | \chi_{\beta} \rangle} < -4 t
\quad {\rm or} \quad
\frac{\langle \chi_{\beta} | \hat{\mathcal{H} } | \chi_{\beta} \rangle}{\langle   \chi_{\beta} | \chi_{\beta} \rangle} < -2 t
\, .
\label{eq:bound}
\end{equation}
Eq.~\eqref{eq:bound} implies that the ground state energy of a single tetrahedron is lower than $-2t$. However, the ground state energy of a single tetrahedron is exactly $-2t$, which implies that $E_g$ cannot be lower than $-4t$.

For completeness, let us close by showing explicitly that the single tetrahedron GS energy is $-2t$. 
If we consider a system consisting of four fully connected sites, labelled $1,2,3,4$, with one holon and three spins, we 
can define the wave functions 
\begin{eqnarray}
    | \Psi^{\rm tet}_{1 \sigma} \rangle &=& \frac{1}{\sqrt{3}} \hat{c}^\dagger_{1 \sigma}\left( \hat{d}^\dagger_{2 3} + \hat{d}^\dagger_{3 4} + \hat{d}^\dagger_{4 2}\right) | 0 \rangle
    \nonumber \\
    | \Psi^{\rm tet}_{2 \sigma} \rangle &=& \frac{1}{\sqrt{3}} \hat{c}^\dagger_{2 \sigma}\left( \hat{d}^\dagger_{1 3} + \hat{d}^\dagger_{3 4} + \hat{d}^\dagger_{4 1}\right) | 0 \rangle 
    \nonumber \\
    | \Psi^{\rm tet}_{3 \sigma} \rangle &=& \frac{1}{\sqrt{3}} \hat{c}^\dagger_{3 \sigma}\left( \hat{d}^\dagger_{1 2} + \hat{d}^\dagger_{2 4} + \hat{d}^\dagger_{4 1}\right) | 0 \rangle 
    \, , \label{eq:singletetWF}
\end{eqnarray}
with $\hat{d}^\dagger_{ij} = \frac{1}{\sqrt{2}} (\hat{c}^{\dagger}_{i \uparrow} \hat{c}^{\dagger}_{j \downarrow} - \hat{c}^{\dagger}_{i \downarrow} \hat{c}^{\dagger}_{j \uparrow} )$.
Note that $\hat{d}^\dagger_{ij} = \hat{d}^\dagger_{ji}$. The state $| \Psi^{\rm tet}_{i \sigma} \rangle$ has a static spin $\sigma$ on site $i$, while the holon and a singlet resonate on the remaining three sites. Consequently, the total spin of the state is $\sigma$. Applying the single tetrahedron Hamiltonian to this state, we find that hopping to site $i$ indeed is cancelled, and the states are eigenstates with energy $-2t$. Note that the state where site $4$ has a static spin is a superposition of the three states in Eq.~\eqref{eq:singletetWF}: $| \Psi^{\rm tet}_{4 \sigma} \rangle = \frac{1}{\sqrt{3}} \hat{c}^\dagger_{4 \sigma}\left( \hat{d}^\dagger_{1 2} + \hat{d}^\dagger_{2 3} + \hat{d}^\dagger_{3 1}\right) | 0 \rangle = -(| \Psi^{\rm tet}_{1 \sigma} \rangle + | \Psi^{\rm tet}_{2 \sigma} \rangle + | \Psi^{\rm tet}_{3 \sigma} \rangle)$.
%
%
%------------------------------------------------------

\subsection{Ground state wave function
\label{app:proofsGS0}}
Consider now the resonant valence bond (RVB) state:
\begin{equation}
| \Psi_{\rm RVB} \rangle = \frac{1}{ \sqrt{\kappa N}} \sum_{ {\bm r}} | \phi^{\rm RVB}_{\bm r} \rangle
\, ,
\label{eq:GS}
\end{equation}
with 
\begin{equation}
| \phi^{\rm RVB}_{\bm r} \rangle = \sum_{a_{\bm r} } | \Psi_{a_{\bm r}} \rangle, \quad \kappa= \langle  \phi^{\rm RVB}_{\bm r} | \phi^{\rm RVB}_{\bm r} \rangle
\, , 
\label{eq:2}
\end{equation}
where $N \to \infty$ is the number of sites, the index $a_{\bm r}$ runs over all dimer coverings with one dimer per tetrahedron on the lattice with the site ${\bm r}$ excluded 
and $| \Psi_{a_{\bm r}} \rangle$ is the corresponding direct product of singlet states on each dimer:
\begin{equation}
%| \Psi^a_{\bm r} 
| \Psi_{a_{\bm r}} 
\rangle = \prod_{\langle i,j \rangle \in a_{\bm r}} \hat{d}^\dagger_{ij} | 0 \rangle
\, .
\label{eq:0}
\end{equation}

Substituting Eq.~\eqref{eq:0} into Eq.~\eqref{eq:2}, and thence into Eq.~\eqref{eq:GS}, we obtain
\begin{equation}
| \Psi_{\rm RVB} \rangle 
= 
\frac{1}{ \sqrt{\kappa N}}  \sum_{{\bm r}, \{  a_{\bm r} \} } | \Psi_{a_{\bm r}} \rangle
=
\frac{1}{ \sqrt{\kappa N}}   \sum_{{\bm r},\{  a_{\bm r} \} } \prod_{\langle i,j \rangle \in a_{\bm r}} \hat{d}^\dagger_{ij} | 0 \rangle
\, .
\label{eq:1}
\end{equation}

From Eq.~\eqref{eq:exp}, we see that 
\begin{equation}
E_{\rm RVB} \equiv \langle  \Psi_{\rm RVB} |  \hat{\mathcal{H}}   | \Psi_{\rm RVB} \rangle  = \frac{1}{\kappa N}
\sum_{\alpha}     \langle \chi^{\rm RVB}_{\alpha} | \hat{\mathcal{H} }^{\alpha} | \chi^{\rm RVB}_{\alpha} \rangle,
\label{eq:exp2}
\end{equation}
with
\begin{equation}
| \chi^{\rm RVB}_{\alpha}  \rangle  = \sum_{{\bm r} \in \alpha} | \phi^{\rm RVB}_{\bm r} \rangle
\, .
\label{eq:chialp}
\end{equation}
According to Eq.~\eqref{eq:2}, 
\begin{equation}
| \phi^{\rm RVB}_{\bm r} \rangle =| \tilde{\Psi}^{{\bm r}_1 {\bm r}_2 }_{{\bm r}} \rangle    +  | \tilde{\Psi}^{{\bm r}_2 {\bm r}_3}_{{\bm r}} \rangle +  | \tilde{\Psi}^{{\bm r}_1 {\bm r}_3}_{{\bm r}} \rangle
\, , 
\label{eq:sumi}
\end{equation}
where ${\bm r}_1,{\bm r}_2,{\bm r}_3$ denote the $3$ sites of the tetrahedron $\alpha$ that are not occupied by the holon (the pair ${\bm r}_i {\bm r}_j$
is equivalent to ${\bm r}_j {\bm r}_i$) and 
\begin{equation}
| \tilde{\Psi}^{{\bm r}_i {\bm r}_j}_{{\bm r}} \rangle = \sum_{a_{\bm r}/ d \in ({\bm r}_i,{\bm r}_j)  } | \Psi_{a_{\bm r}} \rangle
\, , 
\end{equation}
where the sum runs over dimer coverings $| \Psi_{a_{\bm r}} \rangle$ with one holon at site ${\bm r}$ and one singlet on the bond $({\bm r}_i {\bm r}_j)$ with $1 \leq i , j \leq 3$ and $i < j$.
By inserting Eq.~\eqref{eq:sumi} in Eq.~\eqref{eq:chialp}, we obtain:
\begin{equation}
| \chi^{\rm RVB}_{\alpha}  \rangle  = \sum_{{\bm r} \in \alpha, i <j} | \tilde{\Psi}^{{\bm r}_i {\bm r}_j}_{{\bm r}} \rangle 
\, . 
\label{eq:chialp2}
\end{equation}

The sum contains $12$ terms corresponding to the $12$ different ways of allocating one holon (in one of the $4$ sites of the tetrahedron) and one singlet in one of the $3$ pairs that can be formed with the $3$ remaining sites. We will group these $12$ terms in $4$ groups, where each group consists of the sum of three states:
\begin{equation}
| \tilde{\Psi}^{\alpha}_{{\bm r}} \rangle = | \tilde{\Psi}^{{\bm r}_i {\bm r}_j}_{{\bm r}_k} \rangle + | \tilde{\Psi}^{{\bm r}_k {\bm r}_i}_{{\bm r}_j} \rangle + 
| \tilde{\Psi}^{{\bm r}_j {\bm r}_k}_{{\bm r}_i} \rangle
\, .
\label{eq:psitildealphar}
\end{equation}
The coordinates ${\bm r}_i, {\bm r}_j, {\bm r}_k$ denote the $3$ distinct sites of the tetrahedron (one containing the holon and the other two forming a singlet state) that are \emph{different} from the `passive' site ${\bm r}$. We note that when considering only the four sites on the tetrahedron $\alpha$, $| \tilde{\Psi}^{\alpha}_{{\bm r}} \rangle$ correspond exactly to the eigenstates of the single tetrahedron where one site is passive and the three remaining sites contain a holon and singlet resonating.

Next, if we denote the coordinates of the $4$ sites of the tetrahedron $\alpha$ by ${\bm r}_a$ with $1\leq a \leq 4$, we have:
\begin{equation}
| \chi^{\rm RVB}_{\alpha} \rangle = | \tilde{\Psi}^{\alpha}_{{\bm r}_1} \rangle + | \tilde{\Psi}^{\alpha}_{{\bm r}_2} \rangle + | \tilde{\Psi}^{\alpha}_{{\bm r}_3}\rangle + | \tilde{\Psi}^{\alpha}_{{\bm r}_4}\rangle
\, ,
\label{eq:chialp3}
\end{equation}
and 
\begin{eqnarray}
E_{\rm RVB} &=& \frac{1}{\kappa N}
\sum_{\alpha} \langle \chi^{\rm RVB}_{\alpha} | \hat{\mathcal{H}}^{\alpha} | \chi^{\rm RVB}_{\alpha} \rangle \nonumber\\ 
&=& \frac{1}{\kappa N} \sum_{\alpha} \sum_{a, b =1}^4 \langle \tilde{\Psi}^{\alpha}_{{\bm r}_a} | \hat{\mathcal{H} }^{\alpha}
| \tilde{\Psi}^{\alpha}_{{\bm r}_b} \rangle
\, .
\label{eq:exp3}
\end{eqnarray}

Finally, we note that 
\begin{equation}
\hat{\mathcal{H} }^{\alpha}
| \tilde{\Psi}^{\alpha}_{{\bm r}_a} \rangle = - 2 t \, | \tilde{\Psi}^{\alpha}_{{\bm r}_a} \rangle
\, ,
\label{eq:pass}
\end{equation}
implying that
\begin{eqnarray}
E_{\rm RVB} &=& - \frac{2t}{\kappa N} \sum_{\alpha} \sum_{a,b=1}^4 \langle \tilde{\Psi}^{\alpha}_{{\bm r}_a} 
| \tilde{\Psi}^{\alpha}_{{\bm r}_b} \rangle \nonumber \\
&=& - \frac{2t}{\kappa N} \sum_{\alpha} \langle \chi^{\rm RVB}_{\alpha} | \chi^{\rm RVB}_{\alpha} \rangle 
\nonumber \\
&=& 
- \frac{4t}{\kappa N} \sum_{\bm r} \langle \phi^{\rm RVB}_{\bm r}|  \phi^{\rm RVB}_{\bm r} \rangle= - 4 t 
\, .
\label{eq:exp4}  
\end{eqnarray}
Since we have shown that $-4t$ is a strict lower bound for the energy, Eq.~\eqref{eq:exp4} implies that $| \Psi_{\rm RVB} \rangle$ is an exact ground state of $\hat{\mathcal{H} }$. 
Eq.~\eqref{eq:pass} clarifies the name `passive' for the site ${\bm r}$: since $| \tilde{\Psi}^{\alpha}_{\bm r} \rangle$ is an eigenstate of $\hat{\mathcal{H} }^{\alpha}$, the holon never visits that site because of the destructive interference produced by the linear superposition~\eqref{eq:psitildealphar}. 
%
%
%------------------------------------------------------

\subsection{Considerations about the thermodynamic limit
\label{app:proofsThLim}}
The proof holds in the thermodynamic limit, as only in that limit can we accommodate one dimer in each tetrahedron.
Inserting a hole in a finite lattice with an integer number of tetrahedra leads to an unpaired spin that can be associated with a spinon. In this scenario, the dimer coverings are not complete because there is one tetrahedron which does not contain a dimer. We can always choose that tetrahedron to be located at the boundary of the finite lattice and also to be the one containing the unpaired spin.
Since the boundary disappears upon taking the thermodynamic limit, the unpaired spin and the missing dimer become irrelevant in this limit. However, this simple observation has two important implications. Firstly, as expected for an RVB ground state, the spin and the charge of the fermion added to the Mott insulator separate, i.e., the probability of finding the holon at any finite distance from the spinon is zero in the thermodynamic limit. Secondly, on a finite size lattice the spinon must necessarily occupy a passive site of the tetrahedron, implying that the holon will never visit the site occupied by the unpaired spin. 
In other words, the spinon is expected to have infinite mass (static spinon) and the spinon-holon interaction is expected to be repulsive. Both observations are supported by our numerical results. 
%
%
%-----------------------------------------------------

\subsection{Proposed finite-size ground state wave function
\label{app:proofsFinSize}}
From numerics (discussed in Supp.~Mat.~Notes~\ref{app:methods} and~\ref{app:correlations}), we find that the one-hole GS energy $-4t$ can be met in finite size systems and even when the unpaired spin is fixed at a given site; in the latter case the GS is unique. This allows us to conjecture a form for the wave function that is given by the equal amplitude superposition of all dimer-singlet states that have a spin fixed and a delocalised holon, subject to the constraint of no more than one dimer per tetrahedron. 
Remarkably, we find that the conjectured state is identical within numerical accuracy to the ground state found for the 16-site system using ED, and to the 32-site system using DMRG. We also find this to be the case for the two-holes ground state (i.e., the equal amplitude superposition of all dimer-singlet states that have two delocalised holons, subject to the constraint of no more than one dimer per tetrahedron). 
We therefore propose that such GS wave functions are in fact exact for all system sizes. 

Note that any dimer configuration with a fixed spin and a hole (or two holes) necessarily has one less dimer than tetrahedra. Under the condition of no more than one dimer per tetrahedron, this means that there is precisely one tetrahedron without dimers in the configuration. Notably, we find that the proposed GS wave function is invariant to fixing the location of this tetrahedron. 
%
%
%------------------------------------------------------

\subsection{Vanishing wave function in the stoichiometric limit
\label{app:proofsVanishingNoHoles}}
Curiously, we find that our problem does not admit a straightforward stoichiometric limit of vanishing hole density. If we take the equal amplitude superposition of all dimer-singlet states on the pyrochlore lattice, without holons or unpaired spins, the resulting wave function vanishes identically for both the 16-site and the 32-site system. This surprising result -- which to the best of our knowledge had not been pointed out in the literature before -- holds irrespective of whether we impose the constraint of one and only one dimer per tetrahedron (relevant to our RVB liquid state) or we consider all possible dimer configurations in general~\cite{footnote:vanishing2}. 
%~\footnote{We also find that equal amplitude superpositions of dimer-singlet states vanish for small triangular lattices with periodic boundary conditions, but not for square, honeycomb and kagome lattices. We note that proofs of linear independence exist for dimer-singlet states on some of these lattices~\cite{Chayes1989}; however, such proofs rely on open and not periodic boundary conditions.}. 
%
%
%%%%%%%%%%%%%%%%%%%%%%%%%%%%%%%%%%%%%%%%%%%%%%%%%%%%%%

\section{The counter-Nagaoka effect
\label{app:counterNagaoka}}
Itinerant holons in large-$U$ Hubbard models near half-filling are known to influence the spin correlations. For example, Nagaoka's theorem~\cite{Nagaoka1966,Thouless1965} shows that a single hole in the absence of interactions between the spins is capable of inducing a ferromagnetic ground state. The phenomenon is rooted in the fact that spin correlations affect the kinetic energy of the holon. In Nagaoka's case, a ferromagnetic pattern allows for perfect constructive interference of the holon world lines, thus minimising the kinetic energy. The effect is remarkably stable (namely, a finite magnetisation persists) up to 20\% hole doping~\cite{Liang1995,Park2008,Carleo2011,Liu2012} (however, see also Ref.~\onlinecite{Eisenberg2002} for considerations about a possible breakdown mechanism due to phase separation).

More recently, Haerter and Shastry~\cite{Haerter2005} pointed out that the same kinetic mechanism can give rise to strikingly different spin correlations -- namely, weak metallic antiferromagnetism -- if the electronic lattice is frustrated; this occurs, for instance, when the electron hopping amplitude is negative, and the elementary loops on the lattice encompass an odd number of bonds (see also Ref.~\onlinecite{Poilblanc2004} for earlier results hinting at this phenomenon). This counterpart to the Nagaoka effect was further studied in Refs.~\onlinecite{Sposetti2014,Lisandrini2017} and later dubbed the `counter-Nagaoka effect'~\cite{Zhang2018}. 

The nature of the state on the frustrated triangular lattice was investigated numerically~\cite{Sposetti2014,Lisandrini2017, Morera2024} and shown to exhibit correlations that depart significantly from the behaviour expected for the quantum Heisenberg antiferromagnet, towards a more classical nature (although not entirely: a small deviation from classical correlations is observed~\cite{Sposetti2014}). This work also demonstrated that the kinetic energy mechanism was akin to a finite spin interaction term proportional to the hole density (namely, $J_{t} \sim t/N$, where $N$ is the number of lattice sites, in the presence of a single hole). 

The effect has been recently investigated in ladders in Ref.~\onlinecite{Morera2024}; and on a Husimi cactus of triangles (akin locally to a kagome lattice) where it was shown to form a regular valence bond pattern~\cite{Kim2023}. 

To date, studies of the counter-Nagaoka effect have focused on 2D (or quasi-1D) systems, where it is found to lead to magnetically ordered low-energy states. 
In this work, we explore the behaviour on the 3D pyrochlore lattice (see Fig.~\ref{fig:pyrochlore lattice} in the main text), renowned for its strong frustration effects (and our results straightforwardly apply also to a planar arrangement of tetrahedra, also known as the 2D checkerboard lattice). 
%
%
%%%%%%%%%%%%%%%%%%%%%%%%%%%%%%%%%%%%%%%%%%%%%%%%%%%%%%

\section{Models and methods
\label{app:methods}}
\begin{figure}[]
    \centering
        \begin{minipage}[h!]{0.85\columnwidth}
        \centering
        \includegraphics[width=\columnwidth]{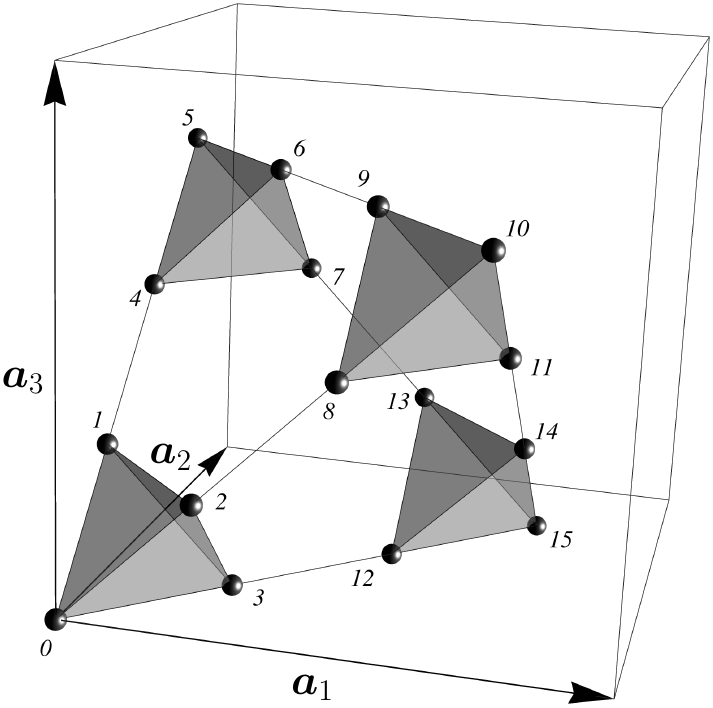}
        \end{minipage}  
    \caption{{\it 16-site pyrochlore system}. This is constructed from a single cubic unit cell with lattice vectors ${\bm a}_1 = (1,0,0)$, ${\bm a}_2 = (0,1,0)$ and ${\bm a}_3 = (0,0,1)$. %The unit cell has 16 sites.
    }\label{fig:lat16}
\end{figure}
\begin{figure}[]
    \centering
        \begin{minipage}[h!]{\columnwidth}
        \centering
        \includegraphics[width=\columnwidth]{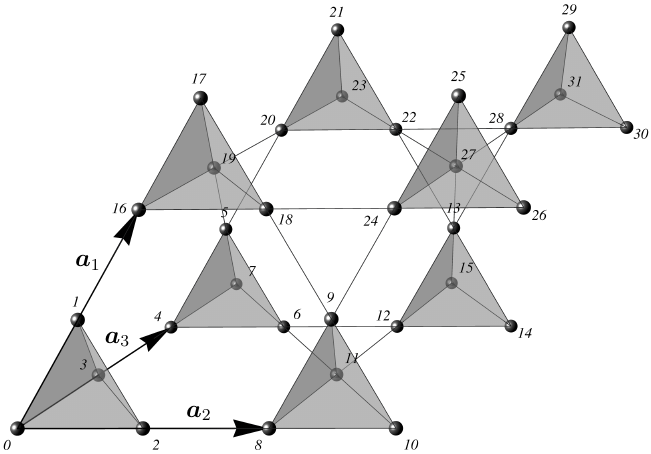}
        \end{minipage}  
    \caption{{\it 32-site pyrochlore system}. This is constructed from a $2\times 2\times 2$ lattice of the ffc unit cell with lattice vectors ${\bm a}_1=(0,1/2,1/2)$, ${\bm a}_2=(1/2,0,1/2)$ and ${\bm a}_3=(1/2,1/2,0)$. Each unit cell has four sites (a single tetrahedron).}\label{fig:lat32}
\end{figure}
In the main text, we expressed the Hamiltonian for our system in terms of constrained electron operators for notational convenience. The constraint can be made explicit by substituting $\hat{c}_{j\sigma} \to \hat{c}_{j\sigma} \hat{P}_{j\sigma}$, where on the right hand side we have the conventional femionic annihilation operator, and $\hat{P}_{i\sigma} = (1-\hat{n}_{i\overline{\sigma}})$ is the projection operator that ensures no double occupancy, with $\hat{n}_{i\sigma}$ being the customary density operator. 
This leads to the Hamiltonian written as 
\begin{equation}
    \hat{\mathcal{H}} 
    = 
    - t\sum_{\langle i,j\rangle \; \sigma} \left[
      \hat{P}_{i\sigma} \hat{c}^\dagger_{i\sigma} \hat{c}_{j\sigma} \hat{P}_{j\sigma} + h.c. 
    \right] 
    + J\sum_{\langle i, j \rangle} \hat{{\bm S}}_i\cdot \hat{{\bm S}}_j
\, ,
%\label{eq:ham_suppmat}
\end{equation}
where once again $\langle i,j\rangle$ denotes nearest-neighbour pairs of sites; $\hat{c}^\dagger_i$ ($\hat{c}_i$) is the fermionic creation (annihilation) operator at site $i$; and $\hat{{\bm S}}_i$ is the corresponding spin operator, $\hat{{\bm S}}_i = \tfrac{1}{2} \sum_{\sigma, \sigma^\prime} \hat{c}^\dagger_{i\sigma} {\bm \sigma}_{\sigma\sigma'} \hat{c}_{i\sigma'}$. 

We study 16- and 32-site systems of the pyrochlore lattice, both with periodic boundary conditions. The 16-site system is a conventional cubic unit cell, illustrated in Fig.~\ref{fig:lat16}. 
The 32-site system is a $2\times 2 \times 2$ spherical cell constructed from the fcc primitive lattice vectors, illustrated in Fig.~\ref{fig:lat32}. 

The 16-site and 32-site systems are chosen as they respect the symmetries of the pyrochlore lattice. Other 
shapes of the unit cell show a tendency to seemingly spurious finite-size effects at very low temperatures in a way akin to partial order-by-disorder effects induced by shape anisotropy. 
%(which is the reason why we did not use 64-site systems to study spin correlations). 
The next possible symmetric system sizes would be the $3\times 3 \times 3$ spherical fcc system, with 108 sites, or a $2\times 2 \times 2$ cubic system with $128$ sites. However, these system sizes are beyond the computational reach of our study. 

For the 16-site system, we perform ED with Lanczos to compute the $800$ smallest eigenvalues and corresponding eigenstates in each magnetisation sector. We ensure we capture all ground states by checking that their number stays the same when the number of computed eigenvalues and eigenstates is doubled. 

\begin{figure}[]
\centering
\includegraphics[width=\columnwidth]{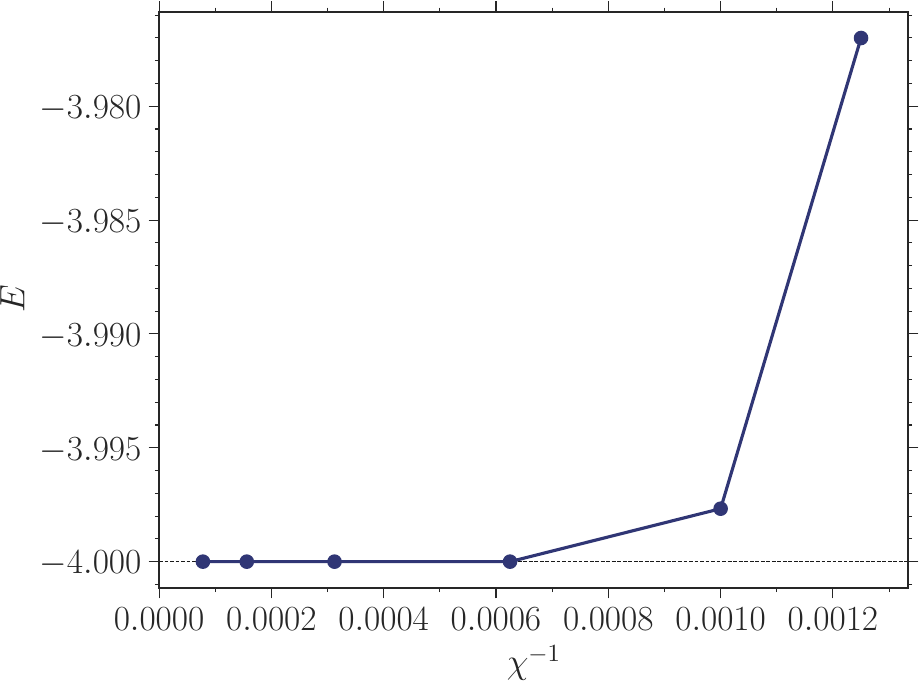}
\caption{{\it Energy as a function of the inverse bond dimension for the 32-site pyrochlore system.} The bond dimensions used are $\chi = \{800,1000,1600,3200,6400,12800\}$.}
\label{fig:chiscaling}
\end{figure}

We use two-site DMRG to study the 32-site system, with an energy convergence criterion of $10^{-10}$ and a maximum bond dimension up to $12800$. This is sufficient to reach convergence, as shown in Fig.~\ref{fig:chiscaling}. Using periodic boundary conditions, we create a one-dimensional snake path of the system that follows the numbering in Fig.~\ref{fig:lat32}. The choice of snake path seems to have little impact on the convergence for our highly entangled 3D system.
The DMRG calculations were performed using the TeNPy Library (version 0.10.0)~\cite{tenpy}. 

A single DMRG calculation gives one ground state. When the ground state is degenerate due to an odd number of spins, we bias each ground state and compute correlations as an average over all ground states. 
In the case of pure hopping ($J=0$, $t=1$), the full GS manifold is found by pinning the unpaired spin by adding an on-site magnetic field. This gives an overcomplete set of ground states, and a linearly independent set of ground states is then found via Gram-Schmidt decomposition.

For the studies of the classical Heisenberg model in Supp.~Mat.~Note~\ref{app:correlations}, we use Monte Carlo (MC) simulations with the Metropolis algorithm~\cite{Newman2011}. The classical structure factors are computed after equilibrating the system down to $\beta = 10^3$. In order to compare quantitatively classical spins with quantum spin $S=1/2$ degrees of freedom, we tune the classical spin length to satisfy $S^2=3/4$. 

As mentioned in the main text, our results hold for lattices of corner sharing tetrahedra and therefore also for the 2D checkerboard lattice. Indeed, the 16-site unit cell of the 2D checkerboard lattice ($4\times2\times2$) is identical to the one for the pyrochlore lattice ($16\times 1\times1\times1$). In our work, we further consider a 24-site ($4\times3\times2$) checkerboard lattice using ED (1 hole) and DMRG (2 holes) as well as 36- ($4\times 3\times3$) and 48-site ($4\times 4\times3$) checkerboard lattices with 1 and 2 holes using DMRG.
%
%
%%%%%%%%%%%%%%%%%%%%%%%%%%%%%%%%%%%%%%%%%%%%%%%%%%%%%%

\section{Doping and correlations
\label{app:correlations}}
In the first instance, we consider the case of non-interacting spins ($t=1$, $J=0$) at zero temperature in the presence of a single hole. For all system sizes, we find that the ground state energy bound is met at $-4t$. For 16 and 32 sites, the GS degeneracy is $18$ and $34$, respectively. 

An overall 2-fold degeneracy can be readily understood from the odd number of spin-$1/2$ degrees of freedom in the system. The residual degeneracy is clearly related to the presence of an unpaired spin. Indeed, we find a unique ground state when the unpaired spin is pinned to a given lattice site, resulting in a set of $16$ and $32$ linearly-dependent states that span the GS manifold. A set of orthonormal GS wave functions can then be found using Gram-Schmidt decomposition, for example, giving the 9-fold and 17-fold degeneracy. 
This is in contrast with the behaviour observed for two holes (see Supp.~Mat.~Note~\ref{app:twoholes}), where there are no unpaired spins and the GS is unique. While the specific degeneracy of the single-hole ground state for arbitrary system sizes remains to be understood, our results suggest that the scaling is at most linear in system size. 

The holon density shows weak repulsive dependence on the distance from the fixed spin, as illustrated in Fig.~\ref{fig:one_hole_corr} in the main text. 
Correspondingly, the expectation value of the hole hopping operator has a suitably inverse dependence, leading to the exact Hamiltonian eigenvalue $-4t$. This is confirmed by computing the expectation value $\langle \sum_{\sigma, j} \hat{c}^\dagger_{i, \sigma} \hat{c}_{j, \sigma} \rangle$, where $j$ runs over all nearest-neighbour sites of $i$. Within numerical accuracy, we find indeed that this expectation value is precisely $4\langle 1- \hat{n}_i \rangle$ for all sites $i$, where $\langle 1- \hat{n}_i \rangle$ is the holon density at site $i$.

We further checked that, within numerical precision, the expectation value of the projector onto the state with maximum angular momentum for each tetrahedron with no holon, 
\begin{equation}
    Q_{\rm tet} = \left\langle 
    \hat{{\bm S}}_{\rm tet}^2
    \left( \hat{{\bm S}}_{\rm tet}^2-2 \right)
    \left(
            \sum_{i\in {\rm tet}}\hat{n}_{i}-3
    \right)
    \right\rangle
    \, , 
\end{equation} 
vanishes identically in the ground state for all tetrahedra. Here, $\hat{{\bm S}}_{\rm tet} = \sum_{i \in {\rm tet}} \hat{{\bm S}}_{i}$ is the total spin of the given tetrahedron; and $\sum_{i\in {\rm tet}}\hat{n}_{i}$ takes on the value $4$ or $3$ depending on whether a holon is absent or present on the tetrahedron, respectively. 

We find within numerical accuracy that the GS wave function (with a fixed spin) from 16-site ED and 32-site DMRG for the pyrochlore lattice, and 16- and 24-site ED
%and 36- and 48-site DMRG 
 for the checkerboard lattice
, is identical to a state constructed as the equal amplitude superposition of dimer-singlet states with a holon and the further constraint of no more than one dimer per tetrahedron. 

\begin{figure}
    \centering      \includegraphics[width=\columnwidth]{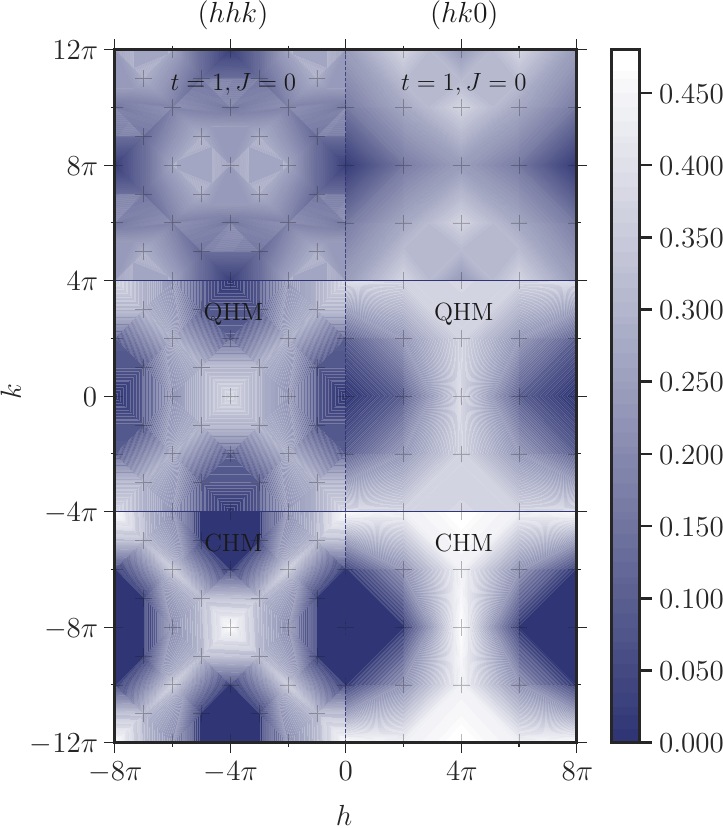}
    \caption{{\it Spin structure factors of the 32-site systems}. Top, doping-induced, $1$ hole ($t=1$ and $J=0$).
    Middle, quantum Heisenberg model (QHM). Bottom, classical Heisenberg model (CHM). Both Heisenberg models have no holes and their correlations are correspondingly scaled by $(N-1)/N$. Accessible momenta are marked by grey pluses.}
    \label{fig:32_doping_QHM_CHM_1hole}
\end{figure}

We compute the spin structure factor of our RVB liquid state, averaged over all ground states of our system for $t=1$ and $J=0$, as $\mathcal{S}({\bm q}) = \sum_{i,j} \langle S_i^z S_j^z \rangle e^{-i{\bm q}\cdot({\bm r}_i - {\bm r}_j)}/N$, where ${\bm r}_i$ is the position of site $i$. This is shown in Fig.~\ref{fig:32_doping_QHM_CHM_1hole} (upper panels) for the 32-site system, obtained using DMRG. There is indeed no evidence of symmetry breaking. While one can recognise features akin to those of the spin-$1/2$ quantum Heisenberg model (obtained using DMRG; see Fig.~\ref{fig:32_doping_QHM_CHM_1hole}, middle panels), the doping-induced correlations in the structure factor of the RVB liquid state in our work are clearly distinct in a way that promises to be measurable in experiments. 
We also compare it for reference to the case of a classical Heisenberg AFM obtained using classical MC simulations (Fig.~\ref{fig:32_doping_QHM_CHM_1hole}, lower panels). Both the quantum and classical pyrochlore Heisenberg AFM are known to be fully frustrated and in a spin liquid GS~\cite{footnote:nematic}. 
%~\footnote{We note however that evidence of possible nematic order has been seen in the quantum case in larger system sizes~\cite{Hagymasi2021, Astrakhantsev2021, Hering2022}.}. 
In the case of the pure Heisenberg models we do not introduce any holes and we scale the resulting correlations by $(N-N_h)/N$, where $N=32$ and $N_h=1$ are the number of sites and holes in the RVB state, respectively. 
%
%
%%%%%%%%%%%%%%%%%%%%%%%%%%%%%%%%%%%%%%%%%%%%%%%%%%%%%%

\section{Two holes
\label{app:twoholes}}
We compute the GS in the presence of two holes and study the GS spin and holon correlations for $t=1$ and $J=0$. We find that the GS is unique with energy $-8t$ in the $S^z = 0$ sector for both the 16-site and 32-site systems, which is exactly twice the GS energy of a single hole. 

We verified that, within numerical accuracy, the GS wave function obtained with ED for 16 sites and DMRG for 32 sites is identical to the equal amplitude superposition of all dimer-singlet states hosting two holons, subject to the constraint of no more than one dimer per tetrahedron. (We also confirmed that this result holds for the 24-site checkerboard lattice.) 

We find that the doping-induced spin correlations when two holes are present (see Fig.~\ref{fig:one_hole_corr} in the main text) are similar to the ones obtained for a single hole. Quantitatively, we compare them using the measure: 
\begin{equation}\label{eq:D}
    D = \frac{1}{N} \sum_{{\bm q}} \left[\vphantom{\sum}
      \mathcal{S}_{\rm 1-hole}({\bm q}) 
      - 
      \mathcal{S}_{\rm 2-holes}({\bm q})
    \right]^2
    \, ,
\end{equation}
and we find values $\lesssim 10^{-4}$ and $\lesssim 10^{-5}$ for the 16-site and 32-site case, respectively, when $\mathcal{S}_{\rm 1-hole}$ and $\mathcal{S}_{\rm 2-holes}$ are suitably rescaled by $(N-N_h)/N$ to account for the different total number of spins. 

\begin{figure}
    \centering
    \includegraphics[width=\columnwidth]{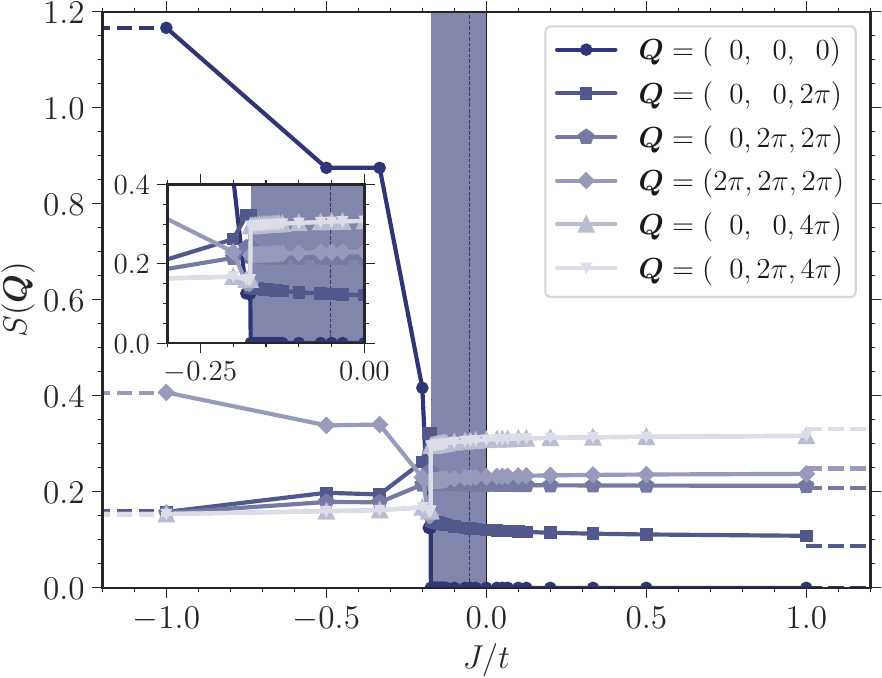}
    %Jovertrange2_new.pdf}
    \caption{{\it Dependence of the ground state spin structure factor on $J/t$} ($t>0$). This is shown for all individual reciprocal lattice points for the $16$-site system with two holes (density $1/8$). Dashed horizontal lines show the $t=0$ limits, for reference. The shaded region denotes correlations that are antiferromagnetic in spite of a ferromagnetic interaction $J$, extending down to $(J/t)_c \simeq -0.173$. For comparison, the dashed vertical line inside the shaded region shows the corresponding extent for hole density $1/16$, $(J/t)_c \simeq -0.052$ (see Fig.~\ref{fig:J/t_one_hole} in the main text).}
    \label{fig:J/t_two_holes}
\end{figure}

Next, we study the effect of having a finite $J$ when two holes are present in the 16-site system. In Fig.~\ref{fig:J/t_two_holes} we show the GS spin structure factor as a function of $J/t$ with $t = 1$. We find that the spin liquid behaviour persists down to a finite negative value of $(J/t)_c \simeq -0.173$ (hole density $12.5$\%). This shows that also in the presence of two holes, quantum spin liquid behaviour can be induced 
%in a ferromagnet on the pyrochlore lattice 
over magnetic ordering 
by frustrated hole doping.

\begin{figure}
    \centering
    \includegraphics[width=\columnwidth]{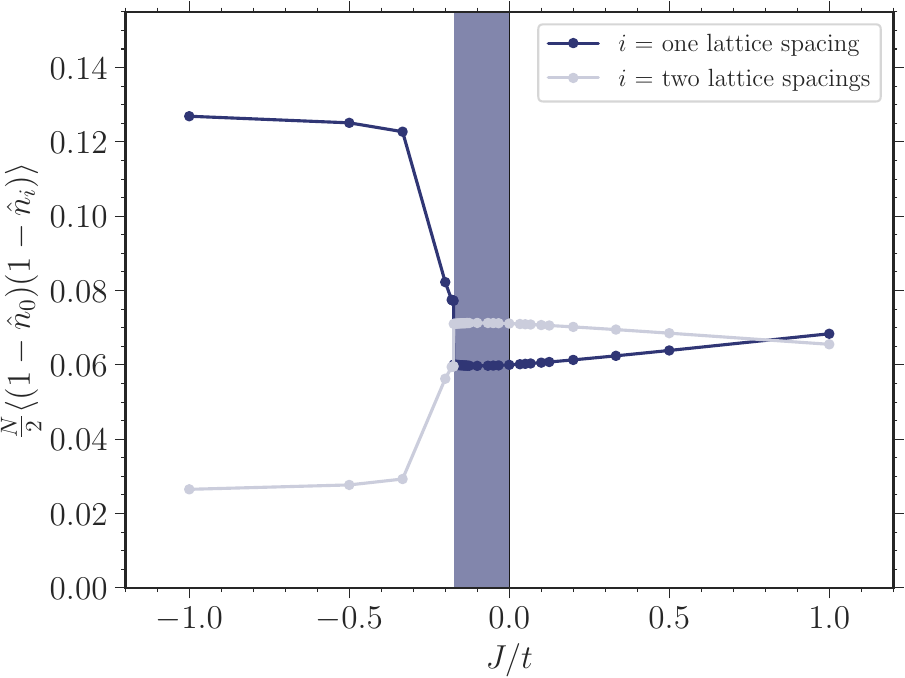}
    %holeholecorr.pdf}
    \caption{{\it Holon-holon correlations as a function of $J/t$}. $\hat{n}_i$ is the spin density operator on site $i$. We measure the correlations between a holon at site 0 and a holon at a site either one or two lattice spacings away. The shaded region shows the range of $J/t$ where the spin correlations are antiferromagnetic in spite of a ferromagnetic interaction $J$.}
    \label{fig:pyrochlore hole-hole correlations J/t}
\end{figure}

Finally, we study the holon-holon correlations. The dependence on distance between the two holes is shown in Fig.~\ref{fig:one_hole_corr} in the main text, for $t=1$ and $J=0$ and system sizes 16 and 32 (and, consistently with the structure of the wave function, it overlaps with the spinon-holon correlation for a single-hole GS). 

For the 16-site system, we also study the behaviour in presence of interactions, as a function of $J/t$, in Fig.~\ref{fig:pyrochlore hole-hole correlations J/t}. Note that because of the small system size, we only have access to three distances (with periodic boundary conditions, two sites in this system are either one or two lattice spacings apart). Additionally the second-nearest neighbour and third-nearest neighbour correlators give the same values, and we therefore only have two independent values of the correlator. 
Within the limitation of the small system size accessed in our study, we observe no sign of holon attraction (rather a small tendency to holon repulsion) in the RVB liquid phase. In the large-$J$ ferromagnetic regime, when the GS becomes partially magnetised, holon attraction is introduced by the holons sharing minority spins~\cite{Zhang2018}. 
%
%
%%%%%%%%%%%%%%%%%%%%%%%%%%%%%%%%%%%%%%%%%%%%%%%%%%%%%%%

\section{Checkerboard
\label{app:checkerboard}}
\begin{figure}[]
\centering
\includegraphics[width=\columnwidth]{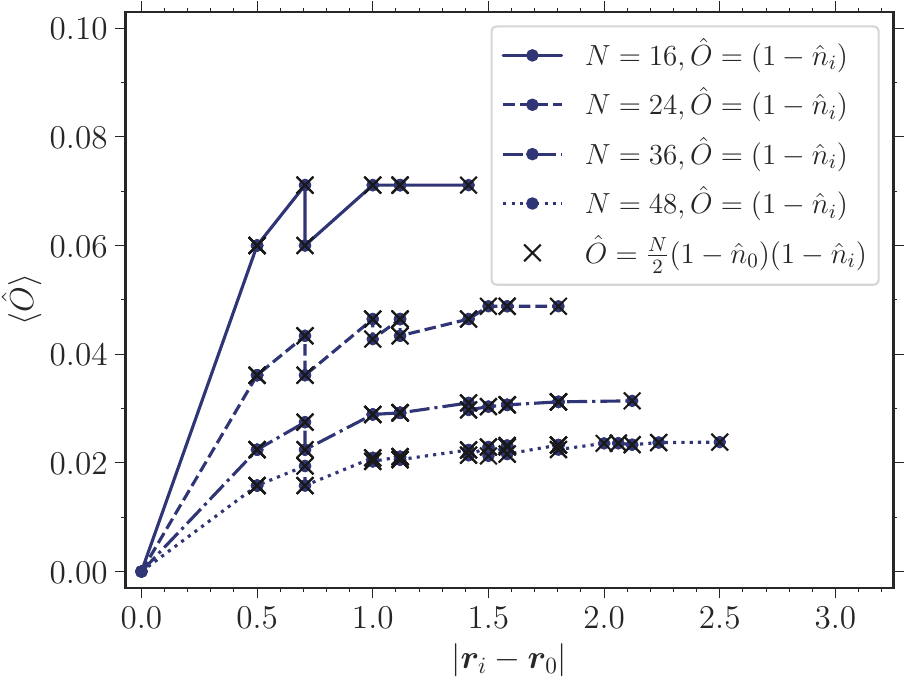}
\caption{{\it Holon correlations on the checkerboard lattice}. The dark-blue dots show the holon density for a checkerboard lattice of 16, 24, 36 and 48 sites, with $t = 1$ and $J = 0$ doped with a single hole, as a function of distance from a pinned spin, indicating spinon-holon deconfinement. Crosses show the holon-holon correlations in equivalent systems with two holes; they are numerically identical to the holon-spinon correlations.}
\label{fig:holonholoncheckerboard}
\end{figure}
In Fig.~\ref{fig:holonholoncheckerboard}, we show the spinon-holon correlations for checkerboard systems of 16, 24, 36, and 48 sites, with one hole and a pinned spin ($t = 1$ and $J = 0$). We include in the figure the holon-holon correlations when two holes are present, for the same system sizes. Once again, we find that the holon-holon correlations exactly match the spinon-holon correlations, consistently with our ground state wave function ansatz. Indeed, as already mentioned earlier, we find that the numerically obtained GS wave function for the 16- and 24-site checkerboard lattice is identical within numerical accuracy to a state constructed as the equal amplitude superposition of dimer-singlet states with one or two delocalised holons (with a fixed spin in the one-holon case) and the further constraint of no more than one dimer per tetrahedron.

\begin{figure}[]
\centering
\includegraphics[width=\columnwidth]{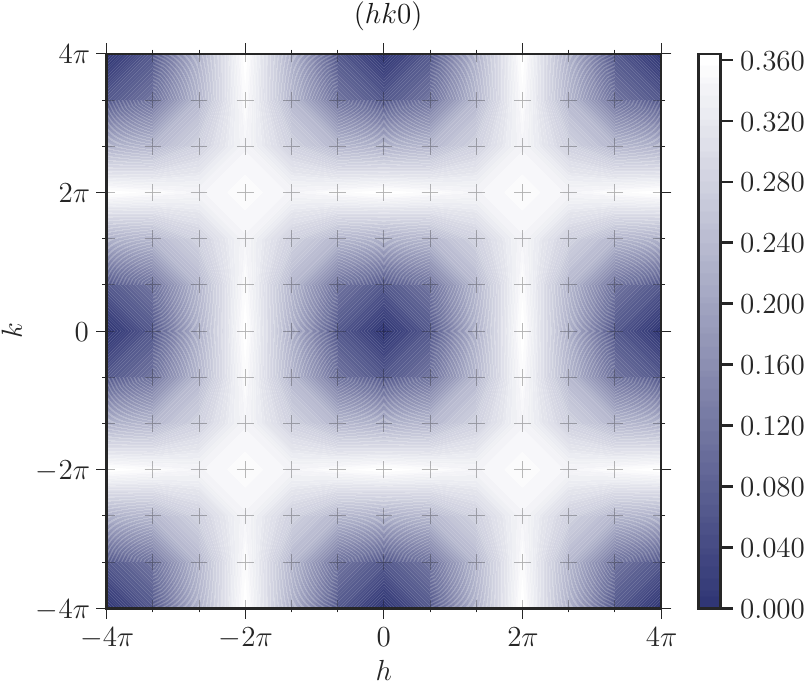}
\caption{{\it Spin structure factor for the 36-site checkerboard lattice.} This is shown for $t = 1$ and $J = 0$, doped with two holes. Accessible momenta are marked by grey pluses.}
\label{fig:structurefactorcheckerboard}
\end{figure}
For a checkerboard system of 36-sites and two holes ($t = 1$ and $J = 0$), we also show in Fig~\ref{fig:structurefactorcheckerboard} the spin structure factor, for reference. 
%
%
%%%%%%%%%%%%%%%%%%%%%%%%%%%%%%%%%%%%%%%%%%%%%%%%%%%%%%%

\section{Experiments
\label{app:expm}}
Among the known pyrochlore materials, the platinum group (Pt, Pd, Rh, Ru, Ir, and Os) of compounds, particularly the iridates and ruthenates, have special interest due to their mobile charge carriers and strong spin-orbit coupling~\cite{Subramanian,WitczakKrempa2012}. 
As discussed in the main text, iridate pyrochlore oxides (RE$_2$Ir$_2$O$_7$, RE = rare earth) are especially promising because -- while being far from a spin-$1/2$ half-filling Hubbard model description -- stoichiometric Ir$^{4+}$ ions have 5d$^5$ shells with effective $J_{\rm eff} = 1/2$ ground state doublets, whereas doping-induced Ir$^{5+}$ have 5d$^4$ shells and $J_{\rm eff} = 0$. In these materials, the iridium ions sit on a lattice of corner-sharing tetrahedra and undergo an antiferromagnetic transition to a long-range ordered state that is characterised by an all-in-all-out (AIAO) arrangement in which the iridium moments point directly into/out of neighbouring tetrahedra~\cite{Tomiyasu2012,Donnerer2016,Jacobsen2020}.
The ordering temperature of the Ir sublattice depends on Ir--O--Ir bond angles and hence is highly sensitive to the size of the RE ion~\cite{Wang2020}. As such, to test our predictions effectively, holes should be introduced while perturbing the IrO$_6$ octahedra as little as possible. In addition, any obfuscating effects of magnetism from the RE sublattice should be avoided. This could be achieved via the synthesis of various doping series in which non-magnetic RE ions (e.g., Y, La, Lu) are successively substituted by an ion of similar size but different charge (e.g., Ca or Sr).
%
%
%------------------------------------------------------

\subsection{Measurements}
Experimentally, the primary indicator of the onset of spin liquid behaviour in the pyrochlore iridates would be the suppression of the AIAO magnetic order on the Ir sublattice as a function of hole doping. In the materials with non-magnetic RE ions, this can be readily observed in the temperature dependence of the magnetic susceptibility. The onset of AIAO order is typically accompanied by a metal-to-insulator transition~\cite{Matsuhira2011}, and resistivity measurements may also be used to track its suppression (with the caveat that a separation of the magnetic and resistive transitions is also possible, as recently observed in Sm$_2$Ir$_2$O$_7$ under applied pressure~\cite{Coak2024}).
Other measurements could be undertaken that provide information on the nature of any observed magnetic ordering, including resonant X-ray scattering, muon spin-relaxation and neutron diffraction~\cite{Wang2020, Jacobsen2020,Disseler2012}. 
Crystal structure determination and careful modelling would also need to be performed to disentangle any effects on the magnetic properties arising from structural changes or from the increase in disorder caused by the introduction of random defects, from those anticipated due to the addition of holes. 

As discussed earlier, some studies of this kind have already been performed, including measurements of polycrystalline samples of 
Y$_{2-x}$Ca$_x$Ir$_2$O$_7$ for 
$x = 0$--0.2 ~\cite{Fukazawa2002,Zhu2014}. 
Using X-ray photoelectron spectroscopy, Ref.~\onlinecite{Zhu2014} finds that the amount of Ir$^{5+}$ in the sample does indeed increase with Ca doping and, at the same time, transport measurements indicate that the conductivity improves. Moreover, a feature in the resistivity that might be associated with the metal-insulator transition seen in other pyrochlore iridates moves to lower temperatures. And temperature-dependent magnetic susceptibility measurements show that the bifurcation of the zero-field and field-cooled traces (which likely arises due to weak ferromagnetic behaviour and is typically associated with the onset of AIAO order in the pyrochlore iridates~\cite{Ishikawa2012}), shifts up in temperature with doping. This last result, however, is in contradiction to Ref.~\onlinecite{Fukazawa2002}, which finds that the bifurcation moves to lower temperatures over the same doping range. That study also finds an enhancement of the conductivity as $x$ increases, but no features related to any metal-insulator transition could be discerned. One drawback of these studies is the polycrystalline nature of the samples. Broken symmetry at the surface and the unknown effects of intergrain boundaries in pressed pellets are known to affect the experimental results in these materials, particularly transport measurements, and tend to obscure what is happening in the bulk~\cite{Coak2024}. So, while a marked effect on the magnetic and transport properties of pyrochlore iridates has already been observed on doping with holes, measurements of high-quality single crystals are necessary to understand these changes and test the predictions discussed here. This requires further investigation into improving synthesis techniques. 
%
%
%------------------------------------------------------

\subsection{Crystal growth}
Polycrystalline samples of pyrochlore iridates (RE$_2$Ir$_2$O$_7$) and ruthenates (RE$_2$Ru$_2$O$_7$) are typically synthesised by conventional solid state sintering between $950^{\degree}$C and $1100^{\degree}$C in air using high purity ($> 99.99$\%) oxide or nitrate chemicals~\cite{Daiki,Haritha}. However, in the case of platinum pyrochlores (RE$_2$Pt$_2$O$_7$), the low decomposition temperature of PtO$_2$ around $450^{\degree}$C makes the conventional ceramic-route synthesis impossible. High-pressure ($30$--$40$~kbar) and high-temperature ($700$--$1200^{\degree}$C) routes have been used to synthesise polycrystalline and small single crystals of platinum pyrochlore materials~\cite{Hoekstra}. The majority of the platinum group compounds belong to the pyrochlore structure with space group Fd$^-$3m (no.277). Since these compounds decompose at high temperature due to high vapour pressure, single crystals cannot be grown using any melt techniques, such as floating-zone or Bridgman. Instead, successful growth of small crystals have been reported via flux methods using KF and CsCl fluxes at $1100^{\degree}$C~\cite{Jasmine,Kristina} (see also our own sample in Fig.~\ref{fig:pyrochlore lattice} in the main text). Recently, high-temperature hydrothermal methods have been employed to synthesise several ruthenate single crystals using KOH as mineraliser~\cite{Bhakti}, and this could be extended to the iridate family.

While the pyrochlore iridates are a promising testbed for the theoretical predictions discussed here, some deviations from the ideal $J = 1/2$ Heisenberg model should be noted. Firstly, the Ir$^{4+}$ ions in RE$_2$Ir$_2$O$_7$ are known to be Ising-like, with the spins preferring to point along the local [111] directions, i.e., into or out of the tetrahedra. It is this anisotropy, coupled with the AFM interactions, that gives rise to the AIAO ordered state at low temperatures and any reduction in the Ising-like nature of the spins (e.g., by substitution of a RE ion with a smaller radius) leads to a suppression of the transition temperature. Secondly, the Ir ions in these materials also exhibit a degree of mixing of the $J_{\rm eff} = 1/2$ state with the $J_{\rm eff} = 3/2$ states due to the local trigonal crystal field (although this is found to be minimised by replacement of the RE ion with some other species). The effects of these deviations from the ideal model are yet to be established, but these materials remain a promising avenue for exploration in this context.  
%
%
%%%%%%%%%%%%%%%%%%%%%%%%%%%%%%%%%%%%%%%%%%%%%%%%%%%%%%%

%\bibliography{references}
\putbib
\end{bibunit}

\end{document}